\documentclass[%
 reprint,
 amsmath,amssymb,
 aps,
pra,
]{revtex4-2}
\usepackage{graphicx}
\usepackage{dcolumn}
\usepackage{bm}

\usepackage{amsmath,amssymb,amsthm,epsfig,color,empheq,graphicx,graphics,algorithm,algorithmic,braket,dsfont,balance}
\usepackage{multirow}
\usepackage[colorlinks=true, linkcolor=blue, citecolor=blue, urlcolor=blue, pdfborder={0 0 0}]{hyperref}

\DeclareMathOperator{\diag}{dg}

\newtheorem{assumption}{Assumption}

\newtheorem{lemma}{Lemma}

\newtheorem{theorem}{Theorem}

\newtheorem{remark}{Remark}

\newcommand \bzero{\mathbf{0}}
\newcommand \bone{\mathbf{1}}

\newcommand \bb{\mathbf{b}}
\newcommand \bc{\mathbf{c}}

\newcommand \be{\mathbf{e}}
\newcommand \bef{\mathbf{f}} 
\newcommand \bg{\mathbf{g}}

\newcommand \bp{\mathbf{p}}

\newcommand \bs{\mathbf{s}}

\newcommand \bx{\mathbf{x}}

\newcommand \bz{\mathbf{z}}
\newcommand \bA{\mathbf{A}}

\newcommand \bC{\mathbf{C}}

\newcommand \bE{\mathbf{E}}
\newcommand \bF{\mathbf{F}}
\newcommand \bG{\mathbf{G}}
\newcommand \bH{\mathbf{H}}
\newcommand \bI{\mathbf{I}}

\newcommand \bR{\mathbf{R}}
\newcommand \bS{\mathbf{S}}

\newcommand \bU{\mathbf{U}}

\newcommand \bW{\mathbf{W}}

\newcommand \bY{\mathbf{Y}}
\newcommand \bZ{\mathbf{Z}}

\newcommand \bdelta{\boldsymbol{\delta}}

\newcommand \btheta{\boldsymbol{\theta}}

\newcommand \blambda{\boldsymbol{\lambda}}

\newcommand \bphi{\boldsymbol{\phi}}

\newcommand \bpsi{\boldsymbol{\psi}}

\newcommand \mcE{\mathcal{E}}

\newcommand \mcG{\mathcal{G}}
\newcommand \mcH{\mathcal{H}}

\newcommand \mcL{\mathcal{L}}

\newcommand \mcP{\mathcal{P}}

\newcommand \mcV{\mathcal{V}}

\newcommand \tbtheta{\tilde{\boldsymbol{\theta}}}

\newcommand \tblambda{\tilde{\boldsymbol{\lambda}}}

\newcommand \hbp{\hat{\mathbf{p}}}

\begin{document}

\preprint{APS/123-QED}

\title{Variational Quantum Eigensolver with Constraints (VQEC):\\
Solving Constrained Optimization Problems via VQE}

\author{Thinh Viet Le}
\email{le272@purdue.edu}
\author{Vassilis Kekatos}
\email{kekatos@purdue.edu}
\affiliation{%
 Elmore Family School of Electrical and Computer Engineering, Purdue University, West Lafayette, IN 47906, USA
}%




\date{\today}

\begin{abstract}
Variational quantum approaches have shown great promise in finding near-optimal solutions to computationally challenging tasks. Nonetheless, enforcing constraints in a disciplined fashion has been largely unexplored. To address this gap, this work proposes a hybrid quantum-classical algorithmic paradigm termed VQEC that extends the celebrated VQE to handle optimization with constraints. As with the standard VQE, the vector of optimization variables is captured by the state of a variational quantum circuit (VQC). To deal with constraints, VQEC optimizes a Lagrangian function classically over both the VQC parameters as well as the dual variables associated with constraints. To comply with the quantum setup, variables are updated via a perturbed primal-dual method leveraging the parameter shift rule. Among a wide gamut of potential applications, we showcase how VQEC can approximately solve quadratically-constrained binary optimization (QCBO) problems, find stochastic binary policies satisfying quadratic constraints on the average and in probability, and solve large-scale linear programs (LP) over the probability simplex. Under an assumption on the error for the VQC to approximate an arbitrary probability mass function (PMF), we provide bounds on the optimality gap attained by a VQC. Numerical tests on a quantum simulator investigate the effect of various parameters and corroborate that VQEC can generate high-quality solutions.
\end{abstract}

\maketitle


\section{Introduction }

Quantum computing could be a disruptive technology in dealing with challenging computational tasks. Seminal works have developed quantum computing algorithms to tackle problems, such as integer factorization~\cite{shor1999}, searching in databases~\cite{grover1996}, solving systems of linear equations~\cite{harrow2009,childs2017}, and various machine learning tasks~\cite{biamonte2017,ciliberto2018}, with polynomial or exponential speed-ups over their classical computing alternatives. However, these algorithms are assumed to operate on fault-tolerant quantum computers, which are projected not to be available in the near future. Recent research and development efforts focus on devising algorithms that are of relevance to practical applications on contemporary, qubit-limited, low-circuit depth, and noisy quantum hardware, often referred to as \emph{Noisy Intermediate-Scale Quantum (NISQ)}~\cite{preskill2018}. Variational quantum approaches (VQAs) exploit parameterized quantum circuits (VQCs) of limited depth and reduced number of qubits, and become the leading candidates to showcase quantum advantage in the NISQ context~\cite{cerezo2021,simeone2022}.

VQE is one of the most well-studied variational quantum approaches. Given a sequence of parameterized quantum gates, VQE aims at seeking the eigenvector corresponding to the minimum eigenvalue (energy) of an exponentially large Hermitian matrix representing a quantum observable~\cite{peruzzo2014}. While VQE has been successfully utilized as a heuristic to find near-optimal solutions for quadratic unconstrained binary optimization (QUBO) problems~\cite{diez2021,barkoutsos2020,harwood2021,GambellaSimonetto,turtletaub2020}, quadratic problems oftentimes come with constraints. A constrained binary problem can be converted to an unconstrained one upon penalizing constraint violations by adding them to the objective function. Nonetheless, the weights associated with each penalty term are non-trivial to select unless the constraints are of specific forms, such as Boolean functions or linear equalities~\cite{Lucas,Ohzeki}. Another VQA that has been particularly successful for binary optimization is the quantum approximate optimization algorithm (QAOA)~\cite{farhi2014}. The QAOA is a special case of VQE that uses a problem-dependent VQC or ansatz. 

To incorporate constraints, references~\cite{Hadfield19,Hadfield17,wang2020} adapt the mixer Hamiltonian of the QAOA's ansatz to ensure that the target quantum state remains within the feasible subspace. This strategy is also studied on quantum annealing, the analog counterpart of QAOA~\cite{HenSarandy,HenSpedalieri}. However, confining the mixer Hamiltonian applies only to a single linear equality constraint and requires a larger number of additional gates~\cite{wang2020}. Reference~\cite{Ronagh} considers binary quadratic programs with linear constraints again, by combining the quantum adiabatic approach with the classical branch-and-bound method. Reference~\cite{GambellaSimonetto} proposes a VQA to minimize an objective expressed as a sum of a quadratic function over binary variables and a convex function over continuous variables. Binary and quadratic variables are set to be equal through linear equality constraints. The problem is solved using the alternating direction method of multipliers (ADMM). Each ADMM iteration entails solving the convex subproblem over the continuous variables using standard convex optimization techniques on a classical computer, and the QUBO subproblem via VQE/QAOA. Nonetheless, solving a VQE or QAOA to optimality for each ADMM iteration may be computationally demanding. In~\cite{shirai2023}, a VQA method is proposed to improve the chance of acquiring feasible solutions for constrained combinatorial problems through a greedy post-processing method, which nonetheless is only applicable to linear constraints. To deal with linear and semi-definite programs by VQA, reference~\cite{westerheim2024} transforms inequality constraints into equality constraints by means of introducing slack variables. Again, the penalty method is employed to modify the equality-constrained optimization to the unconstrained one. Solving the resultant unconstrained problem offers a lower bound on the optimal value of the original constrained problem. In practice, however, the penalty parameter should be chosen carefully, as unreasonably large values could lead to ill-conditioning. To find a suitable value that avoids numerical issues and gives a strict bound at the same time, the unconstrained problem has to be solved for a sequence of penalty parameters.

Lagrangian duality offers a more systematic way of handling optimization problems with constraints. In the context of VQA, the vector of VQC parameters $\btheta$ constitutes the primal optimization variables, and the vector of Lagrange multipliers $\blambda$ corresponds to dual variables. Albeit the primal problem is non-convex, the dual problem is known to be always convex~\cite{BoVa04}. References~\cite{karimi2017,gupta2023} deal with quadratic constrained binary optimization (QCBO) using \emph{dual decomposition}, a variation of subgradient descent that aims at solving the dual problem. Reference~\cite{patel2021} engages dual decomposition too to solve semidefinite programs using variational quantum optimization. However, each update of dual variables involves a complete run of quantum annealing~\cite{karimi2017}, or solving a VQE to optimality~\cite{gupta2023,patel2021}, either of which can be computationally formidable.


In contrast, the primal-dual method used in~\cite{eisen19,gupta2021} to train neural networks under constrained optimization problems is more suitable, since each primal step only requires updating the primal variable inexactly. The convergence of the primal-dual method is guaranteed under strict settings, such as the Lagrangian function being strictly convex and strictly concave~\cite{uzawa1958}. However, by updating the primal and dual variables at appropriate perturbed points, reference~\cite{kallio1994,kallio1999} ensures the sequence of the primal-dual pairs converging to the optimal point without putting strict assumptions on the Lagrangian function. Another variant of the primal-dual method with perturbations is used in~\cite{korpelevich1976}, which is often referred to as the extragradient method (EGM). EGM has been expanded recently to the stochastic setting~\cite{lu2021}. Nonetheless, in the VQC context, EGM increases substantially the number of VQC compilations.

Although our approach is presented for solving optimization problems with constraints, it can be expanded to other setups, such as when a VQC is used as a machine learning (ML) model. This idea has attracted sizable research interest recently; see e.g.,~\cite{benedetti2019,schuld2021}. A popular choice of the loss function for quantum ML is the minimization of expectations of quantum observables concerning a quantum state prepared by the VQC. Although there is no quantum equivalent to automatic differentiation, references~\cite{mitarai2018,schuld2019} derive an analytical formula for the gradient of typical loss functions. However, estimating the gradient through quantum measurements is always subject to noise. Studies~\cite{harrow2021,sweke2020} show that estimating gradients using a finite number of measurements engenders an unbiased estimator and can facilitate stochastic gradient descent. In light of interpreting VQCs as ML models, it is also of crucial importance to explore VQCs from the perspective of learning models for constrained setups.

\subsection{Contributions}
Acknowledging the gap in incorporating constraints into VQAs and its relevance to optimization and ML tasks, the contribution of this work is on four fronts:

\emph{c1)} Section~\ref{sec:solve} puts forth a novel algorithmic framework for handling optimization problems with constraints via a VQA, termed the \emph{variational quantum eigensolver with constraints} (VQEC). This framework applies to problems where cost and constraint functions can be captured as quantum observables over general, exponentially large, Hermitian matrices. As with VQE, rather than solving the problem over the original, exponentially large decision variable, VQEC adopts a hybrid quantum-classical approach. A quantum circuit parameterized over fewer parameters stored in vector $\btheta$ measures the observables, and a classical computer updates $\btheta$ iteratively. To incorporate constraints in a disciplined fashion, VQEC implements a primal-dual method, wherein the classical computer updates not only $\btheta$, but also the vector of Lagrange multipliers associated with the constraints. For improved convergence properties, a perturbed variant termed the perturbed primal-dual (PPD) method is adapted to the quantum setup by capitalizing on the parameter shift rule. Interestingly, if cost and constraint observables can be measured simultaneously, VQEC requires approximately the same quantum computations as VQE.

\emph{c2)} Section~\ref{sec:applications} subsequently exemplifies how VQEC applies to problems with \emph{diagonal observables} (observables defined over diagonal Hermitian matrices), and shows that a wide gamut of optimization tasks can be formulated as such. The list includes problems such as constrained quadratic binary optimization (QCBO), designing stochastic policies over binary-valued vectors that satisfy constraints on the average or in probability as chance constraints, learning large-scale PMFs, and solving large-scale LPs over the probability simplex. Such problems abound in diverse application domains, including reinforcement and machine learning in general, wireless communications, portfolio optimization, and optimal resource allocation.

\emph{c3)} Section~\ref{sec:analysis} provides analytical bounds on the optimality gap experienced when a problem with diagonal observables is solved in its variational form over $\btheta$ rather than its original form over an exponentially large variable. Under an assumption resembling the universal approximation theorem for deep neural networks, the optimality gap is shown to depend on the approximation within which a VQC can approximate any PMF as well as the sensitivity of the original problem to perturbations in the constraints. 

\emph{c4)} Section~\ref{sec:tests} numerically investigates the performance of VQEC on problems with diagonal observables. The tests demonstrate its convergence, reasonable performance even with finitely many quantum measurements, sensitivity to VQC depth, and ability to provide high-quality solutions to binary programs with constraints and large-scale LPs over the probability simplex. 

Section~\ref{sec:conclusions} concludes this work, and sketches exciting ongoing and future research directions for VQEC.


This Introduction closes with a quick note on \emph{notation:} Symbol $(.)^\top$ denotes transposition; symbol $(.)^\mcH$ stands for Hermitian transposition. Matrices (column vectors) are denoted by upper-(lower-) case boldface letters; scalars are denoted by lower-case letters. Operator $\diag(\bx)$ defines a diagonal matrix with vector $\bx$ on its main diagonal. $\mathbb{E}$ is the expectation operator. Calligraphic symbols are reserved for sets. 
\section{Proposed Algorithm}\label{sec:solve}
This section develops a general algorithmic framework for incorporating constraints into variational quantum optimization. This broadens the applicability of VQE to cope with constrained programs over binary variables or large-scale continuous variables. We set the stage and introduce notation with a brief summary of VQE. 

Given a $2^n\times 2^n$ Hermitian matrix $\bH_0$, VQE aims at finding the eigenvector associated with the smallest eigenvalue of $\bH_0$~\cite{peruzzo2014}. From Rayleigh's quotient, this eigenvector coincides with the minimizer of 
\begin{equation}\label{eq:vqe}
\min_{\bx:\bx^\mcH\bx=1}\bx^\mcH\bH_0\bx.    
\end{equation}
Because the dimension $N=2^n$ of $\bx$ is exponentially large, VQE models $\bx$ via the state $\ket{\bx}$ of a variational quantum circuit operating on $n$ qubits. Given a prespecified ansatz, this quantum circuit is parameterized by a parameter vector $\btheta\in\mathbb{R}^P$, and its output state is denoted by $\ket{\bx(\btheta)}$. Rather than solving \eqref{eq:vqe}, VQE solves the ensuing parameterized eigenproblem over $\btheta$:
\begin{equation}\label{eq:vqe2}
\min_{\btheta}~\braket{\bx(\btheta)|\bH_0|\bx(\btheta)}.    
\end{equation}
The VQE operates in a hybrid quantum/classical computing fashion. A quantum computer samples from $\ket{\bx(\btheta)}$, and measures the observable
\begin{equation}\label{eq:F0}
F_0(\btheta):=\braket{\bx(\btheta)|\bH_0|\bx(\btheta)}    
\end{equation}
and possibly its gradient $\nabla_{\btheta}F_0(\btheta)$. A classical computer subsequently optimizes $F_0(\btheta)$ with respect to $\btheta$ using standard optimization techniques, such as coordinate descent and stochastic gradient descent~\cite{schuld2021,sweke2020}.

The cardinal question is how to extend VQE to handle optimization problems with constraints. We consider the prototypical constrained problem
\begin{align}\label{eq:qcboF}
P_{\theta}^*:=\min_{\btheta}~&~F_0(\btheta)\\ 
\mathrm{s.to}~&~F_m(\btheta)\leq 0,\quad m=1:M.\nonumber
\end{align}
Similarly to the cost function in \eqref{eq:vqe2}, constraint functions should be expressible as quantum observables 
\[F_m(\btheta):=\braket{\bx(\btheta)|\bH_m|\bx(\btheta)},\quad m=1:M\]
defined by Hermitian matrices $\bH_m$. Problem~\eqref{eq:qcboF} is non-convex as quantum observables are known to be trigonometric functions of $\btheta$; see~\cite{schuld2021} and references therein. 

Lagrange duality constitutes a systematic way of dealing with constrained optimization problems. Towards solving~\eqref{eq:qcboF}, let $\lambda_m$ be the Lagrange multiplier or dual variable associated with inequality constraint $F_m(\btheta)\leq 0$. Stack all Lagrange multipliers in vector $\blambda\in\mathbb{R}^M$. To simplify notation, let us also define the constant $\lambda_0=1$. Then, the Lagrangian function of~\eqref{eq:qcboF} can be expressed as
\begin{equation}\label{eq:LagrangianTheta}
\mcL_{\theta}(\btheta;\blambda):= \sum_{m=0}^M\lambda_m F_m(\btheta).
\end{equation}
The associated dual function is defined as 
\begin{equation*}
D_{\theta}(\blambda):=\min_{\btheta}~\mcL_{\theta}(\btheta;\blambda)
\end{equation*}
and the corresponding dual problem aims at maximizing
\begin{equation}\label{eq:dual0}
D_{\theta}^*:=\max_{\blambda\succeq \bzero}~D_{\theta}(\blambda).
\end{equation}
The dual problem is convex regardless if the primal problem is convex or not. Moreover, weak duality asserts that $D_{\theta}^*\leq P_{\theta}^*$.

In~\cite{gupta2023}, we attempted solving~\eqref{eq:dual0} using the iterative method of \emph{dual decomposition}. Given an estimate $\blambda^{t}$ of the optimal $\blambda^*$ at the beginning of iteration $t$, iteration $t$ updates the dual variables in two steps:
\begin{subequations}\label{eq:dd}
    \begin{align}
    \btheta^{t}&\in\arg\min_{\btheta}~\mcL_{\theta}(\btheta;\blambda^{t})\label{eq:dd:p}\\
    \lambda_m^{t+1}&:=\left[\lambda_m^{t}+\mu^{t}_{\lambda}F_m(\btheta^{t})\right]_+,~~ m=1:M\label{eq:dd:d}
    \end{align}
\end{subequations}
where $\mu_{\lambda}>0$ is a step size and $[x]_+=\max\{x,0\}$ projects dual variables to the non-negative reals. Dual decomposition is known to be a projected subgradient ascent method to maximize $D_{\theta}(\blambda)$. This is because the constraint function value $F_m(\btheta^{t})$ belongs to the subdifferential of $D_{\theta}(\blambda)$ with respect to $\lambda_m$ evaluated at $\blambda^{t}$. Obviously, the primal update step in~\eqref{eq:dd:p} constitutes a standard (unconstrained) VQE task using the observable $\mcL_{\theta}(\btheta;\blambda^{t})= F_0(\btheta)+\sum_{m=1}^M \lambda_m^{t} F_m(\btheta)$ rather than $F_0(\btheta)$ alone. Unfortunately, solving a VQE task to optimality per iteration of the dual decomposition method is computationally impractical.

To alleviate this limitation, we propose switching from dual decomposition to the so-termed \emph{primal-dual (PD) method}; see~\cite{uzawa1958,zabotin1988}. The latter seeks a \emph{saddle point} of the Lagrangian function over primal/dual variables as
\begin{equation}\label{eq:saddle}
\max_{\blambda\succeq \bzero}\min_{\btheta}~\mcL_{\theta}(\btheta;\blambda).
\end{equation}
A saddle point is a pair $(\btheta^*,\blambda^*)\in \mathbb{R}^P\times \mathbb{R}_+^M$ of primal-dual vectors satisfying
\begin{equation*}
\mcL_{\theta}(\btheta^*;\blambda)\leq \mcL_{\theta}(\btheta^*;\blambda^*) \leq \mcL_{\theta}(\btheta;\blambda^*)
\end{equation*}
for all $(\btheta,\blambda)\in \mathbb{R}^P\times \mathbb{R}_+^M$. For problems where strong duality holds, a pair of primal-dual vectors is optimal if and only if it is a saddle point of the Lagrangian~\cite[p.~239]{BoVa04}. Moreover, the optimal value of the Lagrangian function equals the optimal values of the primal and dual problems. How about the non-convex problem over $\btheta$ in \eqref{eq:qcboF}? Although the latter may not feature strong duality, Section~\ref{sec:analysis} bounds the distance between $\mcL_{\theta}(\btheta^*;\blambda^*)$ and the optimal cost of the original, non-parameterized constrained optimization problem. This motivates solving \eqref{eq:qcboF} using saddle point-seeking algorithms, such as the primal-dual method proposed next.

Instead of a fully-fledged VQE, the PD method updates primal variables by taking a gradient descent step on $\mcL_{\theta}(\btheta;\blambda)$ with respect to $\btheta$, evaluated at $(\btheta^t,\blambda^t)$, that is
\[\btheta^{t+1}:=\btheta^{t}-\mu^{t}_{\theta} \nabla_{\btheta} \mcL_{\theta}(\btheta^{t};\blambda^{t}).\]
The required gradient follows from~\eqref{eq:LagrangianTheta} as
\begin{equation}\label{eq:gradLagrangian}
\nabla_{\btheta}\mcL_{\theta}(\btheta^{t};\blambda^{t})= \sum_{m=0}^M \lambda_m^{t} \nabla_{\btheta}F_m(\btheta^{t})
\end{equation}
with $\lambda_0^{t}=1$ for all $t$. Analogously, the PD method updates $\blambda$ by taking a gradient ascent step on $\mcL_{\theta}(\btheta;\blambda)$ with respect to $\blambda$, evaluated again at $(\btheta^t,\blambda^t)$. The partial derivative of $\mcL_{\theta}(\btheta;\blambda)$ with respect to $\lambda_m$ is $F_m(\btheta)$. 

Putting the two steps together yields a single iteration of the PD method:
\begin{subequations}\label{eq:pdm}
\begin{align}
\btheta^{t+1}&:=\btheta^{t}-\mu^{t}_{\theta} \sum_{m=0}^M \lambda_m^{t} \nabla_{\btheta}F_m(\btheta^{t})\label{eq:pdm:p}\\
\lambda_m^{t+1}&:=[\lambda_m^{t}+\mu^{t}_{\lambda} F_m(\btheta^{t})]_+,~~ m=1:M\label{eq:pdm:d}
\end{align}   
\end{subequations}
where $\mu_{\theta}^t$ and $\mu_{\lambda}^t$ are positive step sizes. We next elaborate on the practical implementation of \eqref{eq:pdm} and comment on the related convergence guarantees.

\emph{Dual update step \eqref{eq:pdm:d}.} To update the vector of multipliers, we need to compute the quantum observables $\{F_m(\btheta^{t})\}_{m=1}^M$. As with VQE and other VQAs, quantum observables cannot be computed exactly, but only estimated through measurements. More specifically, the quantum circuit is parameterized by $\btheta^t$ and is run $S$ times. The number $S$ will be referred to as the number of \emph{measurement shots}. Given these $S$ independent runs, the quantum observable can be estimated classically or quantumly, within accuracy $\epsilon$ if $S$ scales as $O(\epsilon^{-2})$; see~\cite[p.~181]{schuld2021}. For the optimization programs considered in this work and delineated later in Section~\ref{sec:applications}, all $M+1$ quantum observables involved in \eqref{eq:qcboF} are estimated classically and in parallel using the same $S$ measurement shots.

\begin{figure}[t]
\centering
\includegraphics[width=1\linewidth]{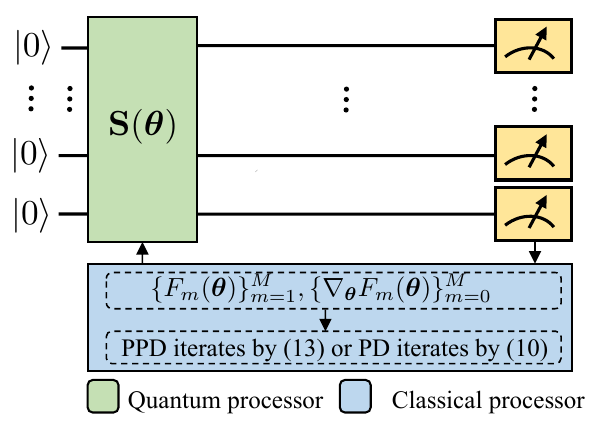}
\caption{Coordination between a quantum and a classical computer while running the PD or PPD method to solve a constrained optimization problem through a variational quantum approach. The VQC is encoded by unitary matrix $\bS(\btheta)$. The proposed method features minimal computational overhead over the standard, unconstrained VQE. If all involved quantum observables $\{F_m(\btheta)\}_{m=0}^M$ can be measured simultaneously, the overhead is insignificant and is confined only to the classical computer.}
\label{fig:PD_scheme}
\end{figure}

\emph{Primal update step \eqref{eq:pdm:p}.} The gradients $\{\nabla_{\btheta}F_m(\btheta^{t})\}_{m=0}^M$ are first estimated with the aid of the quantum computer. The primal variables are then updated using the classical computer. The required gradients can be computed using the \emph{parameter shift rule}~\cite{mitarai2018,schuld2019}. If the ansatz takes the form $\bU(\btheta)=\prod_{i=p}^P \exp(-j\theta_p\bG_p)$, where each $\bG_p$ is a single-qubit Hermitian generator with two distinctive eigenvalues $\pm r$, each partial derivative of $F_m(\btheta)$ with respect to $\theta_p$ can be computed exactly as the difference of two function evaluations at shifted values of $\btheta$:
\begin{equation*}
\frac{\partial F_m(\btheta^t)}{\partial \theta_p}=r\left(F_m(\btheta^t+\tfrac{\pi}{4r}\be_p)-F_m(\btheta^t-\tfrac{\pi}{4r}\be_p)\right)
\end{equation*}
where $\be_p$ is the $p$-th column of the identity $\bI_P$. For example, if $\bG_i$ are the generator matrices corresponding to Pauli rotations $\frac{1}{2} \{\rho_x,\rho_y, \rho_z\}$, then $r=\frac{1}{2}$ and the angle shift is $\frac{\pi}{2}$. As with $F_m(\btheta^t)$ in the dual update step, the quantum observables $F_m(\btheta^t\pm \tfrac{\pi}{4r}\be_p)$ can only be estimated through measurements. Overall, the quantum circuit has to be compiled for $2P$ different values of $\btheta$, two for each partial derivative. For each one of these $2P$ compilations, the quantum circuit is executed $S$ times to estimate $F_m(\btheta^t\pm\tfrac{\pi}{4r}\be_p)$. For the problems considered later in Section~\ref{sec:applications}, quantum observables are estimated classically and in parallel for $m=0,\ldots,M$. Heed that if $\lambda_m^t=0$, there is no need to measure $\nabla_{\btheta}F_m(\btheta^t)$.

Overall, each PD iteration entails compiling the quantum circuit for $2P+1$ different values of $\btheta$ and executing the circuit $(2P+1)S$ times.

The convergence of the PD method has been widely studied for convex problems. For example, under a constant step-size rule a subsequence of the sequence $\{(\btheta_t, \blambda_t)\}$ generated by the PD method is known to converge within a neighborhood of a saddle point of the Lagrangian~\cite{uzawa1958}. For a diminishing step-size rule, a subsequence of the iterates has been shown to converge to a saddle point~\cite{zabotin1988}. To guarantee that the entire sequence of PD updates converges to a saddle point, the function defining the saddle point should be strictly convex in $\btheta$ and strictly concave in $\blambda$~\cite{uzawa1958}, which unfortunately does not hold for the Lagrangian function in general. 

These limitations of the PD method can be resolved using a slight modification, which enjoys convergence of the entire sequence without heavy assumptions on the Lagrangian function~\cite{kallio1994,kallio1999,chang2014}. The so-termed \emph{primal-dual perturbation (PPD) method} updates primal/dual variables upon evaluating gradients $\nabla_{\btheta}\mcL_{\theta}(\btheta;\blambda)$ and $\nabla_{\blambda}\mcL_{\theta}(\btheta;\blambda)$ not at $(\btheta^t,\blambda^t)$, but at a pair of \emph{perturbed} primal/dual variables denoted by $(\tbtheta^t,\tblambda^t)$. In particular, the primal-dual variables are updated as:
\begin{subequations}\label{eq:ppddm0}
\begin{align}
\btheta^{t+1}&:=\btheta^{t}-\mu^{t}_{\theta} \nabla_{\btheta} \mcL_{\theta}(\btheta^{t};\tblambda^{t})\label{eq:ppddm0:p}\\
\lambda_m^{t+1}&:=\left[\lambda_m^{t}+\mu^{t}_{\lambda} F_m(\tbtheta^{t})\right]_+,~~ m=1:M.\label{eq:ppddm0:d}
\end{align}
\end{subequations}
The perturbed variables are updated according to the standard primal-dual method, that is
\begin{subequations}\label{eq:perturbed}
    \begin{align}
    \tbtheta^t &:=\btheta^t-\nu_{\theta} \nabla_{\btheta}\mcL_{\theta}(\btheta^{t};\blambda^{t})\label{eq:perturbed:p}\\
    \tilde{\lambda}_m^t &:= \left[\lambda_m^t+\nu_{\lambda} F_m(\btheta^t)\right]_{+},~~m=1:M\label{eq:perturbed:d}
    \end{align}
\end{subequations}
for positive step sizes $\nu_{\theta}$ and $\nu_{\lambda}$. 

Substituting~\eqref{eq:gradLagrangian} and putting the updates in order, each iteration of PPD involves the next four steps:
\begin{subequations}\label{eq:ppddm}
\begin{align}
\tbtheta^t &:=\btheta^t-\nu_{\theta} \sum_{m=0}^M \lambda_m^{t} \nabla_{\btheta}F_m(\btheta^{t})\label{eq:ppdm:1}\\
\tilde{\lambda}_m^t &:= \left[\lambda_m^t+\nu_{\lambda} F_m(\btheta^t)\right]_{+},~~m=1:M\label{eq:ppdm:2}\\
\btheta^{t+1}&:=\btheta^{t}-\mu^{t}_{\theta} \sum_{m=0}^M \tilde{\lambda}_m^{t} \nabla_{\btheta}F_m(\btheta^{t})\label{eq:ppdm:3}\\
\lambda_m^{t+1}&:=\left[\lambda_m^{t}+\mu^{t}_{\lambda} F_m(\tbtheta^{t})\right]_+,~~ m=1:M\label{eq:ppdm:4}
\end{align}
\end{subequations}
with $\lambda_0^t=\tilde{\lambda}_0^t=1$ for all $t$. The iterations in \eqref{eq:ppddm} constitute the proposed algorithm, termed \emph{variational quantum eigensolver with constraints VQEC}. 

Although the PPD method may seem complicated compared to PD, the additional computations do not incur significant overhead and are primarily run on the classical computer. A simple count of quantum observables to be measured reveals minimal quantum computation overhead. Fortunately, steps~\eqref{eq:ppdm:1} and \eqref{eq:ppdm:3} share the exact same gradients $\nabla_{\btheta}F_m(\btheta^t)$, which can be evaluated using $2P$ compilations of the quantum circuit as in \eqref{eq:pdm:p}. Step~\eqref{eq:ppdm:2} requires compiling the quantum circuit only for $\btheta^t$ as in \eqref{eq:pdm:d}. Step~\eqref{eq:ppdm:4} introduces the sole extra quantum computation as it requires compiling the quantum circuit for $\tbtheta^t$. Overall, the PPD method requires $2P+2$ compilations of the quantum circuit, whereas the PD method needs $2P+1$ compilations. The workflow for implementing the PPD method is illustrated in Fig.~\ref{fig:PD_scheme}. This completes the presentation of VQEC. Additional implementation details are provided in Section~\ref{sec:applications}, as such details pertain to the particular applications considered in this work. A salient feature of VQEC is highlighted next.

\begin{remark}\label{re:salient}
It should be emphasized that so long as cost and constraint observables are compatible (i.e., they can be measured simultaneously), the quantum computations of VQEC do not differ substantially from those of VQE, while the complexity of its classical computations increases only by the number of constraints $M$.
\end{remark}

Some comments on the convergence of PPD are in order. As discussed earlier, the assumptions for the convergence of PD iterates are quite restrictive. PPD iterates enjoy more favorable convergence guarantees. If functions $\{F_m(\btheta)\}_{m=0}^M$ are convex, continuously differentiable, and have Lipchitz continuous gradients, the PPD method with decreasing step sizes $\mu^{t}_{\theta}$ and $\mu^{t}_{\lambda}$ generates a sequence of $\{(\btheta_t,\blambda_t)\}$ converging to an optimal primal/dual pair~\cite{chang2014}. Unfortunately, the parameterized variational problem in \eqref{eq:qcboF} is non-convex, and hence, the mentioned guarantees may not carry over. The convergence analysis of VQEC becomes even more challenging due to its stochastic nature as quantum observables and their gradients are only measured in noise. 

In the context of unconstrained optimization, albeit the convergence of stochastic gradient descent has not been fully analyzed, the method constitutes the spearhead for scaling up deep learning in classical computing. For variational quantum optimization without constraints, preliminary studies on the effect of stochastic gradients can be found in~\cite{harrow2021,sweke2020}. Toward handling constraints, stochastic PD methods have been utilized before in the context of training deep neural networks to satisfy stochastic constraints~\cite{eisen19,gupta2021}. Nevertheless, the convergence of stochastic PD/PPD methods has not been fully established, even in the convex setting. In the convex setting, only saddle point problems featuring particular forms such as bilinear structure, have been analyzed so far. Despite the lack of convergence guarantees, the numerical tests of Section~\ref{sec:tests} study the effect of $S$ and indicate that VQEC iterates do converge to meaningful points and exhibit superior performance over the stochastic PD iterates.


It is worth adding a quick note on another PPD variant, termed the extragradient method~\cite{korpelevich1976}. The perturbed points in EGM are computed as in~\eqref{eq:ppdm:1}--\eqref{eq:ppdm:2}, yet steps~\eqref{eq:ppdm:3}--\eqref{eq:ppdm:4} are altered as 
\begin{align*}
\btheta^{t+1}&:=\tbtheta^{t}-\mu^{t}_{\theta} \sum_{m=0}^M \tilde{\lambda}_m^{t} \nabla_{\btheta}F_m(\tbtheta^{t})\\
\lambda_m^{t+1}&:=\left[\tilde{\lambda}_m^{t}+\mu^{t}_{\lambda} F_m(\tbtheta^{t})\right]_+,~~ m=1:M.
\end{align*}
Although seemingly minor, this modification incurs substantial computational overhead in the quantum setting. This is because measuring $\nabla_{\btheta}F_m(\tbtheta^{t})$ requires $2P$ additional compilations of the quantum circuit. 


\section{Applications}\label{sec:applications}
This section presents prototypical examples of optimization problems with constraints that can be handled by VQEC. Although VQEC applies to quantum observables of the general form $F_m(\btheta)=\braket{\bx(\btheta)|\bH_m|\bx(\btheta)}$, we hereafter focus on observables defined by \emph{diagonal} Hermitian matrices $\bH_m$ for all $m$. This restriction simplifies the process of measuring cost and constraint observables and facilitates the performance analysis of Section~\ref{sec:analysis}. Despite the restriction, quantum observables with diagonal Hermitian matrices can handle a wide gamut of optimization problems, including binary problems with constraints and large-scale LPs over the probability simplex. Quantum observables with non-diagonal Hermitian matrices comprise the subject of our ongoing research. 

Consider a diagonal Hermitian matrix $\bH_m=\diag(\bef_m)$, whose diagonal is defined by vector $\bef_m\in\mathbb{R}^N$. It can be trivially verified that the corresponding quantum observable evaluates as
\begin{equation}\label{eq:dobservable}
F_m(\btheta)=\braket{\bx(\btheta)|\bH_m|\bx(\btheta)}=\bef_m^\top\bp(\btheta)
\end{equation}
where the $k$-th entry of vector $\bp(\btheta)$ is defined as
\begin{equation}\label{eq:ptheta}
p_k(\btheta):=|\braket{k|\bx(\btheta)}|^2,\quad k=0,\ldots,N-1.
\end{equation}
Apparently, observables defined by diagonal Hermitian matrices correspond to inner products. These observables can be measured on the computational basis. They will henceforth referred to as \emph{diagonal observables}. For such observables, the general variational problem in \eqref{eq:qcboF} can be expressed as 
\begin{align}\label{eq:qcboFp}
\min_{\btheta}~&~\bef_0^\top\bp(\btheta)\\ 
\mathrm{s.to}~&~\bef_m^\top\bp(\btheta)\leq 0,\quad m=1:M.\nonumber
\end{align}

Recall that diagonal observables admit an additional neat interpretation. By the properties of $\ket{\bx(\btheta)}$, vector $\bp(\btheta)$ defines a \emph{probability mass function} (PMF). If vector $\bef_m$ carries the $N$ possible values of a discrete random variable distributed according to $\bp(\btheta)$, then $\bef_0^\top\bp(\btheta)$ yields the mean value of this random variable. 

Suppose we measure the VQC state $\ket{\bx(\btheta)}$ in the computational basis. The outcome of this measurement would be binary vectors $\bb\in\{0,1\}^n$ distributed per $\bp(\btheta)$. For each $\bb$, suppose we evaluate a quadratic or other function $f_m(\bb)$. If vector $\bef_m$ carries the evaluations of $f_m$ for all possible values of $\bb$ (i.e., the $k$-th entry of $\bef_m$ is $f_m(\ket{k})$), the diagonal observable provides
\begin{equation}\label{eq:qubos}
F_m(\btheta)=\mathbb{E}_{\btheta}[f_m(\bb)]
\end{equation}
where $\mathbb{E}_{\btheta}$ denotes the expectation operator over the PMF $\bp(\btheta)$ of random variable $\bb$. 

Given the previous interpretations, we next elaborate on what types of problems can be posed as \eqref{eq:qcboFp}. Before doing so, let us recall how VQE handles a famous unconstrained binary optimization problem. 

\subsection{QUBO and MaxCut}\label{subsec:qubo}
The \emph{quadratic unconstrained binary optimization} (QUBO) is one of the problems tackled by VQE. It is originally posed as
\begin{equation}\label{eq:qubo}\tag{QUBO}
\min_{\bb\in\{0,1\}^n}f_0(\bb):=\bb^\top\bA_0\bb +\bb^\top \bc_0+d_0.
\end{equation}
Without harming generality, parameters $(\bA_0,\bc_0,d_0)$ are assumed to be real-valued, and matrix $\bA_0$ is symmetric. QUBO is known to be NP-hard in general, yet VQE-based heuristics have been particularly successful in finding near-optimal solutions~\cite{diez2021,barkoutsos2020,harwood2021,GambellaSimonetto,turtletaub2020}.


We briefly review how VQE is utilized for solving~\eqref{eq:qubo}. If we introduce the $N$-long vector $\bef_0$ whose $k$-th entry is $f_0(\ket{k})$ for $k=0\ldots,N-1$, problem \eqref{eq:qubo} can be equivalently reformulated as
\begin{equation}\label{eq:qubop}
\min_{\bp\in\mcP_c}~\bef_0^\top \bp    
\end{equation}
where $\mcP_c$ is the set of all $N$ canonical vectors in $\mathbb{R}^N$. The minimizer of \eqref{eq:qubop} is the canonical vector corresponding to the smallest entry of $\bef_0$. If \eqref{eq:qubo} has multiple minimizers, problem~\eqref{eq:qubop} will identify one of them. Problem~\eqref {eq:qubop} is as hard as QUBO. 

In pursuit of a computationally more tractable solution, the feasible set of \eqref{eq:qubop} can be relaxed from $\mcP_c$ to its convex hull, that is the \emph{probability simplex} in $\mathbb{R}^N$:
\begin{equation}\label{eq:simplex}
\mcP:=\textrm{conv}(\mcP_c)=\left\{\bp\in\mathbb{R}^N:\bp\geq \bzero,\bp^\top\bone=1\right\}.
\end{equation}
This relaxation yields rise to a linear program (LP)
\begin{equation}\label{eq:qubolp}
\min_{\bp\in\mcP}~\bef_0^\top \bp.
\end{equation}
Problem~\eqref{eq:qubolp} is equivalent to \eqref{eq:qubo} despite the relaxation. This is easy to see since the minimizer of the LP coincides with one of the corners of $\mcP$, or convex combinations thereof. 

Despite being an LP, problem~\eqref{eq:qubolp} is still computationally challenging as $\bp$ is exponentially large. Rather than solving~\eqref{eq:qubolp}, VQE parameterizes $\bp$ through $\ket{\bx(\btheta)}$ as in \eqref{eq:ptheta}, and solves QUBO in the variational form
\begin{equation}\label{eq:qubolptheta}
\min_{\btheta}~\bef_0^\top \bp(\btheta).
\end{equation}
Therefore, QUBO can be posed as an instance of the VQE task in \eqref{eq:vqe2} with a diagonal observable.

\emph{MaxCut} is an instance of QUBO and an NP-hard problem~\cite{karp2010}, for which VQE/QAOA have been successful in finding candidate solutions~\cite{turtletaub2020,crooks2018,zhou2020}. Given an undirected graph $\mcG=(\mcV,\mcE)$ over vertex set $\mcV=\{1,2,..,n\}$ and edge set $\mcE=\{(i,j): i,j\in\mcV\}$ of edges weighted by non-negative $w_{ij}$, MaxCut aims at partitioning $\mcV$ into two subsets so that the size of the \emph{cut} between the subsets is maximized. A cut is a set of edges spanning across two subsets of vertices. The size of a cut is the sum of its edge weights. For instance, in VLSI circuit design, each circuit component is modeled by a vertex. The preference to connect two components in the same or different layers of the circuit is captured by the indicated weight of the edge between them~\cite{dwave2022}.

Let spin variable $s_i\in\{\pm 1\}$ indicate the partition vertex $i$ is assigned to. If vector $\bs\in\{\pm 1\}^n$ collects all spin variables, the cut size defined by assignment $\bs$ is
\begin{equation*}
\frac{1}{4}\sum_{(i,j)\in\mcE}w_{ij}(1-s_is_j).
\end{equation*}
Obviously, edge $(i,j)$ contributes unity to the cut only when $s_i$ and $s_j$ have different signs. If symmetric matrix $\bW$ stores the edge weights as $\bW_{ij}=w_{ij}$, MaxCut is equivalent to minimizing $\bs^\top\bW\bs$. The latter can be written as a QUBO upon converting spin to binary variables via the transformation $b_i=(1-s_i)/2$ for $i=1,\ldots,n$. Having reviewed QUBO and MaxCut, we next embark on incorporating constraints into them.

\subsection{Stochastic QCBO and Learning PMFs}\label{subsec:sqcbo}
Consider the \emph{quadratic constrained binary optimization} (QCBO) problem:
\begin{align}\label{eq:qcbo}
\min_{\bb\in\{0,1\}^n}~&~f_0(\bb)\\ 
\mathrm{s.to}~&~f_m(\bb)\leq 0,\quad m=1:M\nonumber
\end{align}
where $f_m(\bb):=\bb^\top\bA_m\bb +\bb^\top \bc_m+d_m$ for $m=0,\ldots,M$. As with \eqref{eq:qubo}, parameters $(\bA_m,\bc_m,d_m)$ are real-valued, and matrices $\bA_m$ are symmetric for all $m$. Being a generalization of QUBO, QCBO is also NP-hard. A motivating example of a QCBO is discussed next and more can be found in~\cite{wang2010,wang2014,wang2016}.

Albeit MaxCut has been widely studied as an unconstrained problem, constrained versions are of relevance too. In the VLSI design example, the designer may know in advance that specific pairs of components must be on the same or different layers. In that setting, a MaxCut with constraints can minimize the cost of wires used to connect components within and between layers while respecting prior connectivity specifications. Specifications can be encoded in matrix $\bC$ with entries~\cite{wang2014}
\begin{align*}
    \bC_{ij}=
    \begin{cases}
      +1, &\text{if $(i,j)$ are in the same partition}\\
      -1, &\text{if $(i,j)$ are in different partitions}\\
      0, &\text{no prior information on $(i,j)$ or $i=j$}.
    \end{cases}
\end{align*}
We can now define a constrained version of MaxCut~\cite{wang2014}:
\begin{subequations}\label{eq:cMaxCut}
 \begin{align}
\min_{\bs\in\{\pm 1\}^n}~&~\bs^\top\bW\bs\label{eq:cMaxCut:cost}\\
\text{s.to}~&~\bs^\top\bC\bs\geq \sum_{i=1}^n\sum_{j=1}^n|\bC_{ij}|.\label{eq:cMaxCut:con}
\end{align}    
\end{subequations}
Obviously, if a pair $(i,j)$ is correctly assigned to partitions, it contributes $2\bC_{ij}s_is_j=+2$ in the left-hand side of~\eqref{eq:cMaxCut:con}. Otherwise, it contributes $-2$ and the constraint is violated. Upon converting spin to binary variables, problem~\eqref{eq:cMaxCut} can be posed as a QCBO. 

Each specification can also be expressed as a linear constraint $s_i=\bC_{ij}s_j$. However, the formulation in~\eqref{eq:cMaxCut} has a single quadratic constraint instead of multiple linear ones. In fact, the single quadratic constraint can be obtained by squaring and summing up the linear constraints because $\sum_{(i,j):\bC_{ij}\neq 0}(s_i-\bC_{ij}s_j)^2\leq 0$ is equivalent to~\eqref{eq:cMaxCut:con}. In general, linear equality constraints such as $\bE\bb=\bg$ can be handled by a single quadratic constraint $\|\bE\bb-\bg\|_2^2=0$ or $\|\bE\bb-\bg\|_2^2\leq 0$. MaxCut may also come with the balance constraint $-B\leq \bs^\top\bone\leq B$, which ensures that the cardinalities of the two partitions do not differ more than a given constant $B$ from each other.

Given the maturity of mixed-integer linear program (MILP) solvers, quadratic binary programs are oftentimes converted to MILPs. This is possible by introducing an auxiliary variable $z_{ij}$ for each product $b_ib_j$ of binary variables. The constraint $z_{ij}=b_ib_j$ is then handled using McCormick linearization, which requires a few linear constraints involving $(z_{ij},b_i,b_j)$. Nonetheless, this approach can increase the number of constraints and variables by $O(n^2)$. Hence, it may be meaningful to solve quadratic binary problems directly and use quantum computing approaches in particular. Having motivated the need for QCBOs, we next resume with variational quantum approaches for solving them.

Mimicking QUBO, QCBO can be recast as a minimization over the canonical vectors of $\mathbb{R}^N$ as
\begin{align}\label{eq:qcbop}
\min_{\bp\in\mcP_c}~&~\bef_0^\top \bp\\
\mathrm{s.to}~&~\bef_m^\top\bp\leq 0,\quad m=1:M.\nonumber
\end{align}
As with $\bef_0$, each $N$-long vector $\bef_m$ evaluates the $m$-th quadratic function over all possible values of the binary vector $\bb$. The feasible set of \eqref{eq:qcbop} can be subsequently relaxed from $\mcP_c$ to the probability simplex $\mcP$, to yield the large-scale LP
\begin{align}\label{eq:qcbolp}
\min_{\bp\in\mcP}~&~\bef_0^\top \bp\\
\mathrm{s.to}~&~\bef_m^\top\bp\leq 0,\quad m=1:M.\nonumber
\end{align}

Contrary to QUBO, the minimizer of \eqref{eq:qcbolp} may not be at a vertex of $\mcP$. Hence, the problem in \eqref{eq:qcbolp} is \emph{not} equivalent to \eqref{eq:qcbo}. Nonetheless, problem~\eqref{eq:qcbolp} is of interest in its own right as will be explicated shortly. We first explain how~\eqref{eq:qcbolp} can be solved variationally and then discuss possible applications. 

Solving \eqref{eq:qcbolp} classically is technically challenging. Computing the values of vectors $\{\bef_m\}_{m=0}^M$ alone requires $O(NMn^2)$ operations. The size of $\bp$ precludes interior-point methods, while first-order methods would require at least $O(MN)$ operations per iteration only to evaluate the constraint functions. In contrast, a quantum approach could offer a more practical solution as delineated next. Again, variable $\bp$ is substituted by its parameterized form $\bp(\btheta)$, and problem~\eqref{eq:qcbolp} is surrogated by the variational problem in \eqref{eq:qcboFp} with diagonal observables having $\bH_m=\diag(\bef_m)$ for $m=0,\ldots,M$. Consequently, it can be handled by VQEC.

We coin \eqref{eq:qcbolp} and its variational form as the \emph{average QCBO} for the following reason. The VQC state $\ket{\bx(\btheta)}$ can be used as a \emph{sampler} of binary vectors $\bb\in\{0,1\}^n$ drawn from PMF $\bp(\btheta)$. From the viewpoint of~\eqref{eq:qubos}, the sampled binary vectors minimize the average cost $\mathbb{E}_{\btheta}[f_0(\bb)]$ and satisfy constraints in the average sense $\mathbb{E}_{\btheta}[f_m(\bb)]\leq 0$ for $m=1,\ldots,M$. 

The average QCBO can be alternatively interpreted as the task of \emph{learning a PMF}. The PMF $\bp(\btheta)$ is designed to satisfy specifications when applied to given functions. This could be of relevance to machine learning tasks over exponentially large PMFs. Such PMFs arise when dealing with \emph{joint} PMFs over discrete-valued random variables (categorical), and/or probability density functions (PDFs) over continuous random variables that have been finely quantized. Moreover, in certain applications (e.g., reinforcement learning, wireless communications, optimal scheduling), it may be of interest to find a \emph{stochastic policy} to draw binary vectors from, that solve the average QCBO of \eqref{eq:qcboF}. The optimized quantum circuit $\ket{\bx(\btheta)}$ can serve as such policy. The policy can also be used to sample candidate solutions for the deterministic QCBO in~\eqref{eq:qcbo}, yet more elaborate solutions for that follow.

\subsection{QCBO and Chance-Constrained QCBO}\label{subsec:ccqcbo}

So far, vectors $\bef_m$ were assumed to evaluate quadratic functions $f_m(\bb)$ of binary vectors. Nevertheless, functions $f_m(\bb)$ do not have to be quadratic necessarily. Let us see an interesting example. Consider again the QCBO in~\eqref{eq:qcbo} and define functions
\begin{align}\label{eq:gfun}
g_m(\bb)=
\begin{cases}
1,&f_m(\bb)\leq 0\\
0,&f_m(\bb)>0
\end{cases},\quad m=1,\ldots,M.
\end{align}
We next let vectors $\bef_m$ evaluate functions $g_m(\bb)$ rather than $f_m(\bb)$ for $m=1,\ldots,M$, and evaluate the corresponding diagonal observables with Hermitian matrices $\bH_m=\diag(\bef_m)$. It is not hard to verify that the new constraint observables compute the probability
\[F_m(\btheta)=\mathbb{E}_{\btheta}[g_m(\bb)]=\Pr\left(f_m(\bb)\leq 0\right).\]
Vector $\bef_0$ still evaluates the original quadratic $f_0(\bb)$, and so observable $F_0(\btheta)$ remains unchanged.

Thanks to the previous modeling, the constraint
\begin{equation}\label{eq:chance}
F_m(\btheta)=\bef_m^\top\bp(\btheta)\geq 1-\beta
\end{equation}
guarantees that when drawing binary vectors from $\bp(\btheta)$, they satisfy $f_m(\bb)\leq 0$ with probability larger than $1-\beta$. Here $\beta$ is a small positive constant capturing the violation probability. Measuring these observables entails counting the frequency at which each one of the logical statements $(f_m(\bb)\leq 0)$ evaluates as true. Constraint~\eqref{eq:chance} can be made to comply with \eqref{eq:qcboFp} because
\[\bef_m^\top\bp(\btheta)\geq 1-\beta=(1-\beta)\bone^\top\bp(\btheta)\]
is equivalent to
\[((1-\beta)\bone-\bef_m)^\top \bp(\btheta)\leq 0.\]
The latter constraint can take the form $\bef_m^\top\bp(\btheta)\leq 0$ needed in \eqref{eq:qcboFp} with yet another simple change in $\bef_m$'s.

The previous discussion shows that the variational form in \eqref{eq:qcboFp} allows for dealing with \emph{chance-constrained QCBOs}. This allows us to design stochastic policies from which we can draw binary vectors satisfying quadratic constraints with a prescribed probability. Joint chance constraints can be captured too if we define an observable counting the frequency of multiple logical statements being satisfied simultaneously. Designing policies satisfying (joint) chance constraints may be of interest to application domains such as wireless communications. Clearly, setting $\beta=0$ provides a heuristic for dealing with the original \emph{deterministic QCBO} in \eqref{eq:qcbo}.

\subsection{Large-Scale LPs on the Probability Simplex}\label{subsec:lslp}
Lastly, the constrained variational problem in \eqref{eq:qcboFp} can be used to deal with large-scale LPs over the probability simplex. In this case, vectors $\bef_m$'s may bear arbitrary entries, not provided by a particular function. Such large-scale LPs could appear in different application domains, including optimal resource allocation, portfolio optimization, or learning large-scale PMFs from data. In this case, measuring an observable entails VQEC reading out a particular entry of $\{\bef_m\}_{m=0}^M$. The entries of these vectors may not correspond to evaluations of quadratic or other functions. Dealing with such LPs is still computationally challenging on a classical computer, and hence, variational quantum solutions could be welcome.

\subsection{Implementation Details}\label{subsec:implementation}
Some remarks are now due on some implementation details of VQEC. We start with measuring observables. For the average QCBO setting, observables can be computed by running the VQC for a particular $\btheta^t$, measuring its state to get a binary vector $\bb_s$, and evaluating the quadratic function $f_m(\bb_s)$. The process is repeated $S$ times for the same $\btheta$ and observable $F_m(\btheta^t)$ is estimated as
\begin{equation}\label{eq:hFm}
\hat{F}_m(\btheta^t)=\frac{1}{S}\sum_{s=1}^Sf_m(\bb_s).
\end{equation}
For the deterministic and chance-constrained QCBOs, a similar process estimates the frequency at which each constraint is violated across all measurement shots. For the case of large-scale LPs, measuring $F_m(\btheta^t)$ entails reading out the entries of $\bef_m$'s indexed by the sampled $\bb_s$'s and computing the sample mean of these entries. 

Gradients $\{\nabla_{\btheta}F_m(\btheta^t)\}_{m=0}^M$ can be measured similarly thanks to the parameter shift rule. For a particular $\btheta^t$, we need to compute $\hat{F}_m(\btheta^t\pm\tfrac{\pi}{4r}\be_p)$ as discussed earlier, for $p=1,\ldots,P$. Hence, the complete measuring process has to be repeated $2P$ times.

Interestingly, the parameter shift rule applies to \emph{all} problem types identified in this section, i.e., regardless of whether observables evaluate quadratic, binary, or other functions, or simply read out the coefficient vectors of an LP. It is worth stressing that because all observables are diagonal, they can be measured \emph{simultaneously}. In other words, the number of constraints $M$ does not affect the number of VQC compilations or runs of VQEC as all cost/constraint functions and gradients are estimated classically using the same measurements. 

\begin{enumerate}
    \item[\emph{P1)}] The average QCBO of \eqref{eq:qcboF}.
    \item[\emph{P2)}] The deterministic QCBO in \eqref{eq:qcbo} expressed as \eqref{eq:qcboF} with constraints $F_m(\btheta)\geq 1$. Here observables evaluate binary functions by counting the probability of satisfying quadratic constraints.
    \item[\emph{P3)}] Chance-constrained QCBOs expressed as \eqref{eq:qcboF} with constraints $F_m(\btheta)\geq 1-\epsilon$, for small $\epsilon\geq 0$. \emph{P2)} is a special case of \emph{P3)} for $\epsilon=0$.
    \item[\emph{P4)}] Binary optimization problems with cost and constraint functions more general than quadratic, which are nonetheless, easy to compute or measure. 
    \item[\emph{P5)}] Large-scale LPs over the probability simplex as in \eqref{eq:qcbolp}. 
\end{enumerate}

\subsection{Discussion}\label{subsec:discussion}
The previous discussion suggests that depending on how vectors $\bef_m$'s are defined, the formulation in \eqref{eq:qcboFp} can handle a wide variety of optimization problems. Vectors $\bef_m$ essentially define the objective and constraint observables. Vector $\bef_m$ may not necessarily evaluate quadratic functions on the computational basis. Instead, it may evaluate a binary-valued, polynomial, or other analytic function on the computational basis. The presumption is that this function can be efficiently evaluated or measured on a classical (or possibly quantum) computer. Is there any price paid when observables do not correspond to quadratic functions? We do not have an answer to this question but only provide some thoughts. When diagonal observables originate from quadratic functions, the VQC can be tailored to the problem at hand as in QAOA. Non-quadratic diagonal observables, on the other hand, are not amenable to efficient implementations of problem Hamiltonians in the ansatz. Among the applications presented, only the average QCBO seems to be amenable to a QAOA implementation. One may also argue that compared to evaluating logical expressions as in $g_m(\bb)$, quadratic functions $f_m(\bb)$ exhibit continuity, which may help VQEC in more accurately estimating observables or optimizing over the landscape induced by the VQC.

It is worth noting that VQC optimization could suffer from the so-termed \emph{barren plateaus}, where gradients vanish exponentially fast in terms of the VQC size (number of qubits, number of gates, and/or circuit depth)~\cite{mcclean2018}. Vanishing gradients may cause gradient-type algorithms to stall. Although such issues may not be easy to encounter in small-sized examples under which current VQCs are currently tested, they could inhibit VQC's adoption to larger real-world-sized problems. It should be stressed that barren plateaus are a challenge for all VQC-based approaches, not VQEC alone. It is expected that approaches for handling barren plateaus would carry over to VQEC. For example, recent results show that judiciously initializing VQC's parameters could help in avoiding barren plateaus~\cite{Grant2019,park2024}. By transforming VQC into a circuit without fixed entangling gates, reference~\cite{park2024} proves there exist landscapes of VQC parameters, where gradients have large magnitudes regardless of the circuit depth. This result is examined on the hardware efficient ansatz, which is prevalently used in VQC studies due to its suitability for NISQ devices. Albeit finding an initial large gradient does not ensure the algorithm is free from barren plateaus during the training procedure, numerical tests evince that such initialization strategies indeed aid training VQC efficiently~\cite{Grant2019,park2024}.

\section{Performance Analysis}\label{sec:analysis}
This section analyzes the degradation in performance when the optimization problems of Section~\ref{sec:applications} are solved in their quantum variational form of~\eqref{eq:qcboFp} instead of the original form in~\eqref{eq:qcbop}, which is an exponentially large LP over the probability simplex. Collecting the linear inequality coefficient vectors $\{\bef_m\}_{m=1}^M$ as columns of an $N\times M$ matrix $\bF$, the LP in \eqref{eq:qcbolp} can be written as
\begin{align}\label{eq:primal}
P^*=\min_{\bp\in\mcP}~&~\bef_0^\top \bp\\
\mathrm{s.to}~&~\bF^\top\bp\leq \bzero:\quad \blambda.\nonumber
\end{align}
Its dual function can be expressed as
\begin{equation}\label{eq:dual}
D(\blambda)=\min_{\bp\in\mcP}~\mcL(\bp;\blambda)
\end{equation}
where the related Lagrangian function is defined as 
\begin{equation}\label{eq:Lagrangian}
\mcL(\bp;\blambda):=(\bef_0+\bF\blambda)^\top\bp.
\end{equation}
The associated dual problem maximizes the dual function over the dual variables as
\begin{equation}\label{eq:dual*}
D^*=\max_{\blambda\geq \bzero}~D(\blambda).
\end{equation}
If the primal problem is feasible, strong duality holds and yields that $D^*=P^*$.

Let us next consider \eqref{eq:primal} in its variational quantum form, where variable $\bp$ is parameterized by $\btheta$. Rephrasing~\eqref{eq:qcboFp}, the parameterized primal problem reads as  
\begin{align}\label{eq:primalP}
P_{\theta}^*=\min_{\btheta}~&~\bef_0^\top \bp(\btheta)\\
\mathrm{s.to}~&~\bF^\top\bp(\btheta)\leq \bzero:\quad \blambda.\nonumber
\end{align}
The corresponding dual function is
\begin{equation}\label{eq:dualP}
D_{\theta}(\blambda):=\min_{\btheta}~\mcL_{\theta}(\btheta;\blambda)
\end{equation}
with the related Lagrangian function defined as
\begin{equation}\label{eq:LagrangianP}
\mcL_{\theta}(\btheta;\blambda):=(\bef_0+\bF\blambda)^\top\bp(\btheta).
\end{equation}
The parameterized dual problem is expressed as
\begin{equation}\label{eq:dual*P}
D_{\theta}^*=\max_{\blambda\geq \bzero}~D_{\theta}(\blambda).
\end{equation}

The original problem is convex, yet exponentially large. The variational problem is over $\btheta\in\mathbb{R}^P$ with $P\ll N$, yet non-convex. Migrating from the former to the latter lurks two risks. The first risk is that VQEC may not converge, or converge to a local optimum or stationary point. Studying the convergence of (stochastic) PD methods for non-convex problems is challenging, and is not addressed here. Given the success of stochastic gradient-based methods for non-convex problems in deep learning, we only investigate the convergence issues of VQEC numerically in Section~\ref{sec:tests}. The second risk is the possible degradation in optimality, which constitutes the topic of this section. 

Although \eqref{eq:primalP} is non-convex, standard results from duality theory assert that \eqref{eq:dual*P} is a convex problem and that weak duality holds so that $D_{\theta}^*\leq P_{\theta}^*$. The goal of the ensuing analysis is dual: \emph{i)} Characterize the duality gap $P_{\theta}^*-D_{\theta}^*$. If the duality gap of \eqref{eq:primalP} is zero, the saddle point of the Lagrangian function $\mcL_{\theta}$ corresponds to optimal primal/dual solutions for the parameterized problem. Hence, aiming for a saddle point of the Lagrangian is equivalent to seeking the optimal primal/dual variables; and \emph{ii)} Compare $D_{\theta}^*$ with $D^*$. The parameterized problem is apparently a restriction of the original problem, which entails that $P^*_{\theta}\geq P^*$. Because bounding the difference $P^*_{\theta}-P^*$ is challenging, we study the difference $D_{\theta}^*-D^*$ instead. The methods proposed in Section~\ref{sec:solve} to deal with \eqref{eq:qcbolp} target at optimizing $\mcL_{\theta}$ anyway, which at optimality equals $D_{\theta}^*$, that is $D_{\theta}^*=\mcL_{\theta}^*$ by definition of the latter.

Toward the first goal, consider two assumptions.

\begin{assumption}\label{as:convexP}
Consider a VQC whose state $\ket{\bx(\btheta)}$ induces the parameterized PMF vector $\bp(\btheta)$. The set of PMF vectors that can be produced by all admissible $\btheta$'s 
\[\mcP_{\theta}:=\{\bp:~\bp=\bp(\btheta)~\text{for some}~\btheta\}\]
is a convex set.
\end{assumption}

Note that $\mcP_{\theta}$ is a set over $\bp\in\mathbb{R}^N$, and more specifically $\mcP_{\theta}\subseteq\mcP$. It is not a set over $\btheta$.

\begin{assumption}\label{as:strictlyfeasible}
The parameterized primal problem in \eqref{eq:primalP} is strictly feasible, i.e., there exist $\btheta$ and $\bs_{\theta}> \bzero$ for which $\bF^\top\bp(\btheta)=-\bs_{\theta}<\bzero$.    
\end{assumption}

\begin{theorem}\label{th:zdg}
Under Assumptions~\ref{as:convexP} and \ref{as:strictlyfeasible}, the parameterized problem in \eqref{eq:primalP} has zero duality gap, that is $D_{\theta}^*=P_{\theta}^*$.
\end{theorem}

Theorem~\ref{th:zdg} provides two sufficient conditions under which the parameterized problem features zero duality gap; the proof of this theorem as well as all other claims can be found in the Appendix. Unfortunately, Assumption~\ref{as:convexP} may not hold for practical VQCs. If it does hold, then not only $D_{\theta}^*=P_{\theta}^*$, but also $P_{\theta}^*=P^*$. To show this, the next lemma is needed. 

\begin{figure}[t]
\centering
\includegraphics[width=0.75\linewidth]{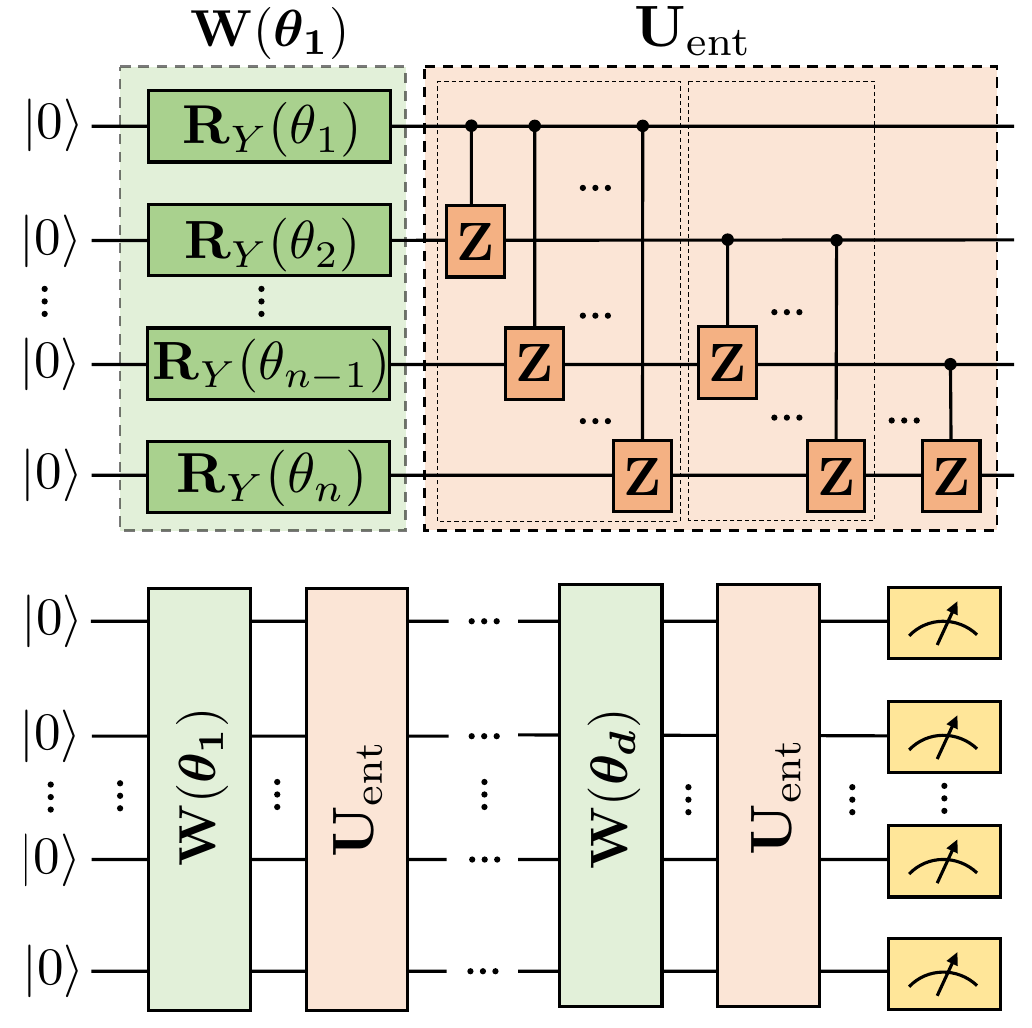}
\caption{The two-local VQC operating on an $n$-qubit system. \emph{Top:} One layer of the two-local VQC consists of the parameterized block $\bW(\btheta_1)$ followed by the full entanglement block $\bU_\text{ent}$. \emph{Bottom:} A $d$-layered two-local VQC.}
\label{fig:two_local}
\end{figure}

\begin{lemma}\label{le:corners}
The variational PMF $\bp(\btheta)$ induced by the two-local VQC with full entanglement depicted in Fig.~\ref{fig:two_local} can capture all corner points of the probability simplex $\mcP$. In other words, for $k=0,\ldots,N-1$, there exists a $\btheta_k$ for which $\bp(\btheta_k)=\ket{k}$.
\end{lemma}

According to Lemma~\ref{le:corners}, the domain $\mcP_{\theta}$ of $\bp(\btheta)$ includes the corners of $\mcP$. If additionally, set $\mcP_{\theta}$ is convex per Assumption~\ref{as:convexP}, then $\mcP_{\theta}=\mcP$ since $\mcP$ is the convex hull of its corners. Consequently, the variational problem is not a restriction anymore, and $P_\theta^*=P^*$. Because Assumption~\ref{as:convexP} is unlikely to be satisfied by practical VQCs, we aim at characterizing the optimality gap under a milder assumption on the VQC.

\begin{assumption}\label{as:epsilon}
For every $\bp\in\mcP$, there exists a $\btheta$ for which $\|\bp-\bp(\btheta)\|_{\infty}\leq \epsilon$ for some $\epsilon>0$.
\end{assumption}


Assumption~\ref{as:epsilon} assumes that the domain of $\bp(\btheta)$ is sufficiently dense so that any PMF vector can be approximated entrywise within accuracy $\epsilon$. This assumption resembles the \emph{universal approximation property} established for and widely used in deep neural networks; see e.g.,~\cite[Sec.~6.4.1]{Goodfellow2016}. Although the observables defined over asymptotically rich VQCs have been shown to be universal function approximators~\cite{schuld2021effect}, the argument may not be trivially extendable to the PMF setup, where a VQC aims to approximate an $N$-long vector rather than a univariate function. An additional assumption on a perturbed version of the original primal LP of \eqref{eq:primal} will be also needed.  

\begin{assumption}\label{as:perturbedprimal}
Consider the linear program
\begin{align}\label{eq:primalperturbed}
\tilde{P}^*=\min_{\bp\in\mcP}~&~\bef_0^\top \bp\\
\mathrm{s.to}~&~\bF^\top\bp\leq -\epsilon L\bone:\quad \blambda\nonumber
\end{align}
where $\epsilon$ has been defined under Assumption~\ref{as:epsilon}, and
\begin{equation}\label{eq:L}
L:=\max_{m=1:M}~\|\bef_m\|_1.
\end{equation}
The assumption is that \eqref{eq:primalperturbed} is strictly feasible. In other words, there exist $\hbp\in\mcP$ and $s_0>0$ satisfying
\begin{equation}\label{eq:strictfeasibleperturbedprimal}
\bF^\top\hbp\leq -\epsilon L\bone - s_0 \bone.
\end{equation}
\end{assumption}

Using Assumptions~\ref{as:strictlyfeasible} and inspired by the proofing procedure of~\cite{eisen19}, the next result bounds the degradation in performance when surrogating the original exponentially-large LP by its variational form. 

\begin{theorem}\label{th:degradation}
Under Assumptions~\ref{as:strictlyfeasible}, \ref{as:epsilon}, and \ref{as:perturbedprimal}, the optimal dual value of the variational quantum problem satisfies
\begin{equation}\label{eq:degradation}
D^*\leq D_{\theta}^*\leq D^*+\epsilon\|\bef_0\|_1+\epsilon L\|\tblambda\|_1
\end{equation}
where $L=\max_{m=1:M}\|\bef_m\|_1$ and $\tblambda$ is the vector of optimal Lagrange multipliers for the perturbed primal problem in \eqref{eq:primalperturbed}. The norm $\|\tblambda\|_1$ can be bounded as 
\begin{equation}\label{eq:degradation2}
\|\tblambda\|_1\leq \frac{\bef_0^\top\hbp - P^*}{s_0}.
\end{equation}
\end{theorem}

Theorem~\ref{th:degradation} predicates that the optimality gap in the dual domain $D_\theta^*-D^*$ is affected by two factors: \emph{i)} The accuracy $\epsilon$ within which the VQC can approximate PMF vectors; and \emph{ii)} The sensitivity of the primal LP to $\epsilon$-perturbations in the cost and constraints. 

\begin{remark}\label{re:equalities}
Assumptions~\ref{as:strictlyfeasible} and \ref{as:perturbedprimal} presume that both the original and the perturbed problems are strictly feasible. Such a requirement automatically excludes problems with equality constraints. Variational problems with equality constraints can be handled by Theorem~\ref{th:degradation} only if equality constraints are relaxed to double-side inequality constraints. Despite this limitation in performance analysis, the algorithms of Section~\ref{sec:solve} remain directly applicable to equality-constrained variational problems.
\end{remark}

\section{Numerical Tests}\label{sec:tests}
VQEC was numerically evaluated under three setups, which are representative of the application examples presented in Section~\ref{sec:applications}:
\renewcommand{\theenumi}{\emph{S\arabic{enumi}})}
\begin{enumerate}
    \item This setup solves the average QCBO of Section~\ref{subsec:sqcbo} on the constrained MaxCut problem. A weighted $n=14$-vertex graph was randomly generated. To include constraints, we randomly sampled $7$ pairs of vertices and added connection specifications so the problem was feasible.
    \item This setup deals with the deterministic QCBO of~\eqref{eq:qcbo} using the formulation of Section~\ref{subsec:ccqcbo}, applied to the same instance as in \emph{S1)}.
    \item This setup solves an LP over the probability simplex like the one in~\eqref{eq:qcbolp} as described in Section~\ref{subsec:lslp}. Given the restrictions of quantum simulators, the problem dimension was $N=256$ with $M=3$ constraints. Vectors $\{\bef_m\}_{m=0}^3$ were stacked into a $256 \times 4$ matrix $\bF$, whose entries were randomly generated from the standard normal distribution.
\end{enumerate}

All numerical tests were performed on quantum simulators from IBM's Qiskit~\cite{Qiskit}. Simulation scripts were written in Python. The two-local VQC from Qiskit was used across tests. As illustrated in Fig.~\ref{fig:two_local}, each layer of the VQC consists of a parameterized block implemented via single-qubit $\bR_Y(\theta_p)$ gates applied to each qubit and a full entanglement block implemented via $CZ$ gates between all pairs of qubits. The number of parameterized blocks of the VQC is referred to as the \emph{circuit depth} $d$. Clearly, the length $P$ of $\btheta$ relates to the problem size $n$ and the circuit depth $d$ as $P=dn$. The PPD/PD iterations were deemed to have converged when $\|\btheta^{t}-\btheta^{t-1}\|_2/\|\btheta^{t-1}\|_2\leq \epsilon$ for $\epsilon = 1\cdot10^{-5}$. The initial primal vector $\btheta^0$ was drawn uniformly within $[0,2\pi]$ using a fixed seed across tests, while $\blambda^0=\bzero$.

Regarding \emph{S1)}, we first compared the convergence properties of VQEC. As noted in Section~\ref{sec:solve}, PPD updates with decreasing step size converge to optimal solutions for convex problems~\cite{kallio1994,kallio1999,chang2014}. Given no analogous result for non-convex problems, we investigated the convergence of VQEC numerically. To eliminate stochasticity due to measurement noise, the initial test utilized the \texttt{statevector\_simulator} quantum simulator from IBM's Qiskit. The two-local VQC used in this test has two full layers and an additional parameterized block yielding a depth of $d=3$. In this case, the test implements the exact PPD/PD methods rather than their stochastic variants. Step sizes followed a time-decreasing rule as $\mu^k_{\theta}=1.5/k$, $\mu^k_{\lambda}=0.1/(t+15)$, while the additional ones for PPD were set to $\nu_{\theta}= \nu_{\lambda}=0.05$. 

\begin{figure}[t]
\centering
\includegraphics[width=1\linewidth]{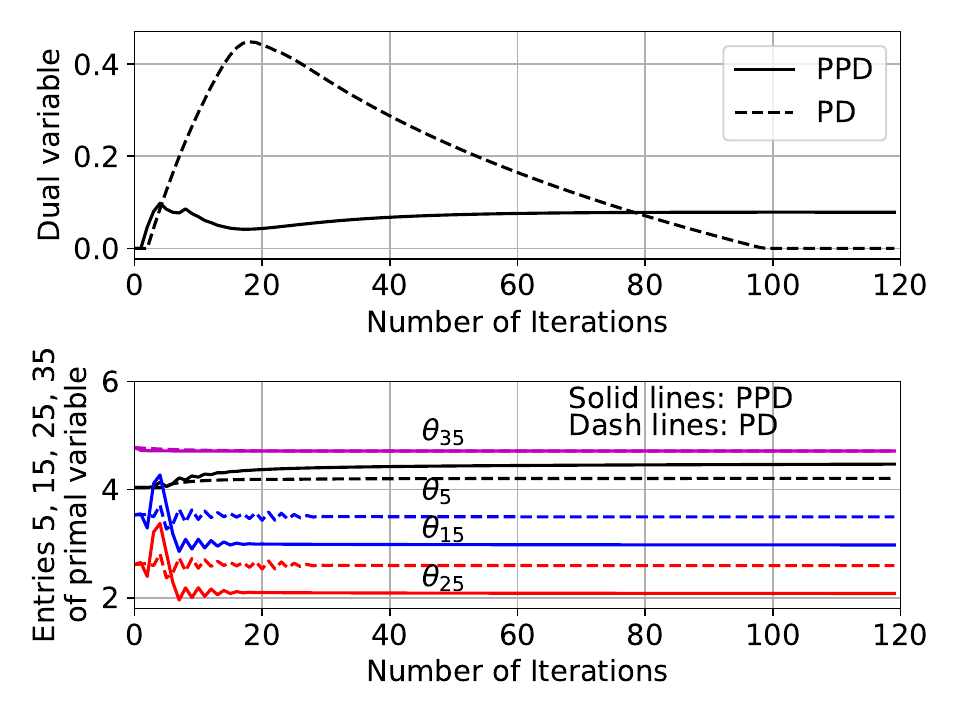}
\caption{Convergence of primal/dual variables for PD and PPD methods over the number of primal/dual iterations under setup \emph{S1)} using a quantum state simulator (no measurement noise). \emph{Top:} Convergence of dual variable $\lambda$. Because the constraint is active, the optimal $\lambda$ is nonzero. \emph{Bottom:} Convergence of entries 5, 15, 25, 35 of the primal variable $\btheta$.}
\label{fig:dual_primal_plainvsperturbed}
\end{figure}

\begin{figure}[t]
\centering
\includegraphics[width=1\linewidth]{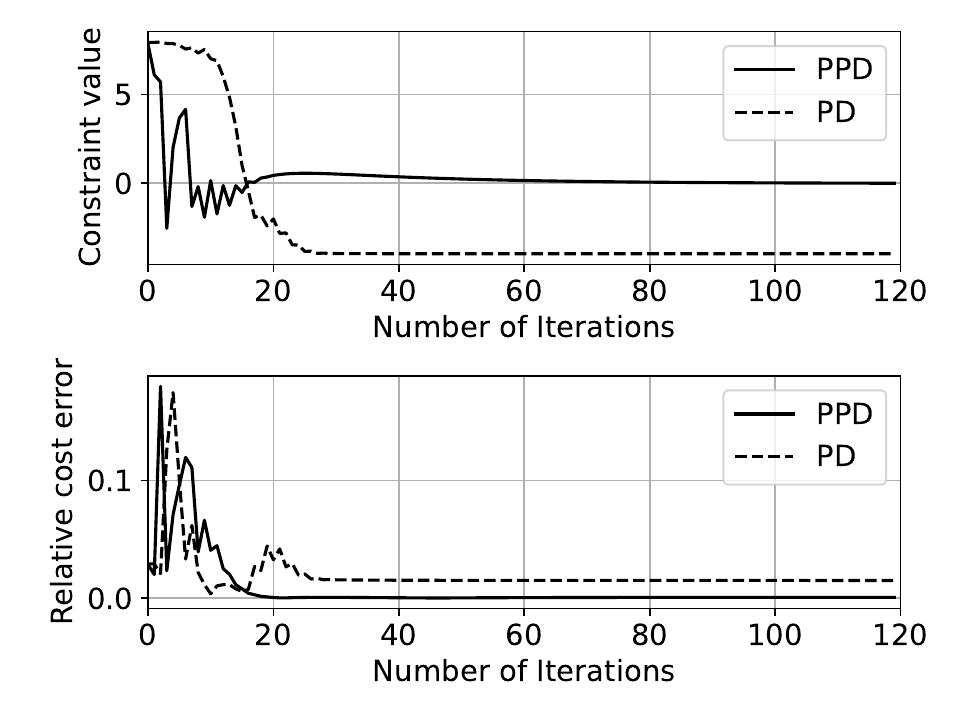}
\caption{Convergence of constraint function value $F_1(\btheta^t)$ (top) and relative cost error $|(P_\theta^t-P^*)/P^*|$ (bottom) for the PD and PPD methods over the number of primal-dual iterations under setup \emph{S1)} using a quantum state simulator (no measurement noise).} 
\label{fig:gval_fval_plainvsperturbed}
\end{figure}

For comparison, we also solved the original average QCBO in \eqref{eq:qcbolp} to optimality using the Gurobi solver under the YALMIP environment~\cite{gurobi,lofberg2004yalmip}. The optimal cost and constraint function values were used as references to verify the feasibility and optimality of the variational solutions found by VQEC. Due to sign invariance, MaxCut solutions come in pairs. The deterministic MaxCut problem has a unique pair of solutions. Nonetheless, the average QCBO returns a PMF vector with four non-zero entries. If we convert these entries to indices of canonical vectors, two of these canonical vectors are optimal and two are infeasible for the original deterministic MaxCut with constraints. Hence, this particular problem instance is non-ideal as the average QCBO does not solve the deterministic QCBO. 

Figure~\ref{fig:dual_primal_plainvsperturbed} depicts the convergence of primal/dual variables using PD and PPD (VQEC). The constraint is active (i.e., satisfied with equality) for this studied instance. Both PD and PPD converged after roughly 500 iterations. However, as shown in Fig.~\ref{fig:dual_primal_plainvsperturbed}, the dual variable of PD converged to 0, whereas the one of PPD converged near 0.1. The VQC parameters for the two methods converged to different values too. Regardless of the converged $\btheta$, how did the trained VQCs perform in terms of feasibility and optimality? Figure~\ref{fig:gval_fval_plainvsperturbed} depicts the convergence of constraint function $F_1(\btheta^t)$ and relative cost error $|(P_\theta^t-P^*)/P^*|$, where $P^*$ is the optimal cost found by Gurobi. Evidently, both the constraint function and the cost error converged to zero using PPD, whereas the ones obtained by PD have greatly deviated from zero. Although PD found a feasible solution, that solution yielded suboptimal cost. On the other hand, PPD found the optimal solution in this case. This test evinces that the employed VQC is capable of finding a $\bp(\btheta)$ that coincides with the optimal solution of~\eqref{eq:qcbolp}, and also that PPD can converge to the saddle point of the Lagrangian function of~\eqref{eq:LagrangianP}. It should be emphasized that PPD is not guaranteed to converge to a global optimum of the non-convex variational problem in general. Nonetheless, given the advantage demonstrated by the previous and similar tests, all subsequent tests use the PPD (rather than the PD) method. 

\begin{figure}[t]
\centering
\includegraphics[width=1\linewidth]{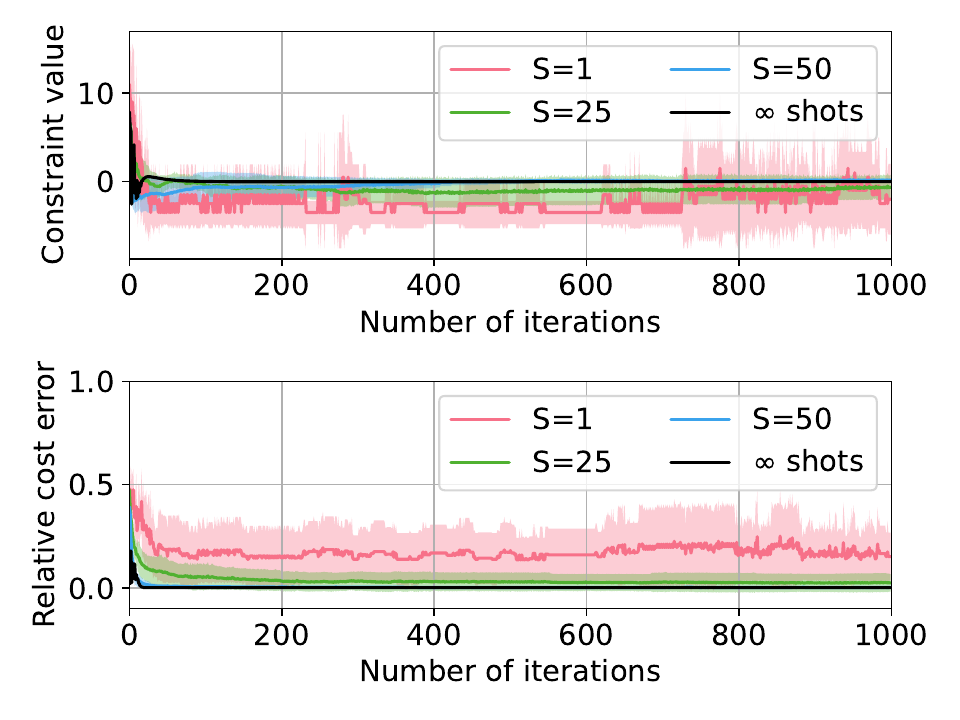}
\caption{Convergence of constraint function value $F_1(\btheta^t)$ (top) and relative cost error $|(P_\theta^t-P^*)/P^*|$ (bottom) for PPD over the number of primal-dual iterations under \emph{S1)} using different values of measurement shots $S$. For $S\in\{1, 25, 50\}$, PPD was repeated 8 times to account for the randomness in iterations. The plots display confidence intervals within one standard deviation around the mean per iteration.}
\label{fig:T1_over_iters}
\end{figure}

The previous test utilized \texttt{statevector\_simulator}, which is equivalent to measuring observables using an infinite number of measurement shots $S$ in~\eqref{eq:hFm}. In practice, observables are measured using a finite $S$ giving rise to \emph{stochastic} PPD updates. To assess the effect of using a finite number of measurement shots, we ran PPD iterates under \emph{S1)} with $S$ taking values in $\{1, 25, 50, \infty\}$. The \texttt{aer\_simulator} was used for finite values of $S$. The circuit depth was set to $d=3$. Figure~\ref{fig:T1_over_iters} shows the convergence of the constraint value, and relative cost error of the proposed method compared to the exact cost value solved by the Gurobi solver. For each finite $S$, PPD iterations were repeated 8 to account for stochasticity. As can be seen from Fig.~\ref{fig:T1_over_iters}, the constraint values and cost errors given by $S=1$ are highly variant and away from zero. For increasing $S$, the constraint value and cost error not only move closer to zero but also exhibit less variance across iteration instantiations. Compared to the state estimator (infinite $S$), PPD with $S=50$ is competent to sample near-optimal solutions under \emph{S1)}. This test demonstrates the improvement in convergence of VQEC for increasing $S$. Nevertheless, the comparison across $S$ may not be fair as a PPD iteration with $S=50$ requires running the VQC 50 times more than a PPD iteration with $S=1$. Recall also that each PPD iteration requires $(2P+2)S$ measurement shots.

\begin{figure}[t]
  \centering
    \includegraphics[width=1\linewidth]{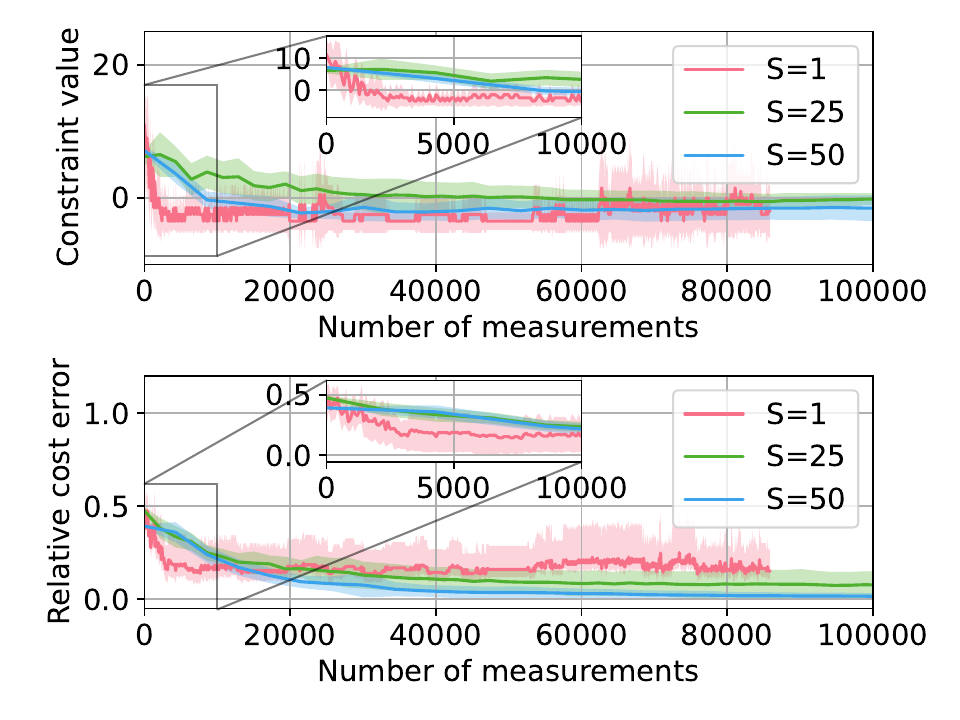}
  \caption{Convergence of constraint function value $F_1(\btheta^t)$ (top) and relative cost error $|(P_\theta^t-P^*)/P^*|$ (bottom) for PPD over the total number of measurement shots while solving \emph{S1)} for different $S$. For $S\in\{1, 25, 50\}$, PPD was repeated 8 times. The plots display confidence intervals within one standard deviation around the mean per iteration.}
  \label{fig:T1_over_meas}
\end{figure}

To study VQEC's convergence over the total VQC runs, Figure~\ref{fig:T1_over_meas} plots the convergence of the constraint value and relative cost error over the total number of measurement shots. As highlighted in the two insets, for the first $10,000$ shots, the constraint value and cost error of the $1$-shot method moved to the zero line faster than the ones of the 25-shot and 50-shot methods did. Eventually, however, the constraint value and the cost error of the 1-shot method remained away from zero, even after $86,000$ shots. On the other hand, the $25$- and $50$-shot iterations approached zero. The solution generated by the $50$-shot method in particular is feasible and its cost error is near-optimal after $100,000$ measurement shots. Given these observations, it may be meaningful to use $S=1$ at early iterations and deliberately increase $S$ as time goes by, to save the total number of measurement shots without compromising optimality. This strategy was also suggested for standard VQE without constraints in~\cite{sweke2020,kubler2020}. Since $S=50$ attained a good trade-off between optimality and total number of shots, it was used for all subsequent tests.

\begin{table}[t]
\centering
\caption{Worst-case probability of obtaining optimal solutions when solving the original QCBO of~\eqref{eq:qcbo} through the variational average QCBO and the variational deterministic QCBO.}
\begin{tabular}{*{4}{p{11mm}}}
\hline\hline
\multicolumn{1}{|c|}{\multirow{2}{*}{\textbf{Variational Problem}}} & \multicolumn{3}{c|}{\textbf{Probability of success}}                               \\ \cline{2-4} 
\multicolumn{1}{|c|}{}                                      & \multicolumn{1}{c|}{$S=1$} & \multicolumn{1}{c|}{$S=25$} & \multicolumn{1}{c|}{$S=50$} \\ \hline\hline
\multicolumn{1}{|l|}{Average QCBO}                              & \multicolumn{1}{l|}{0.0000}   & \multicolumn{1}{l|}{0.5240}   & \multicolumn{1}{l|}{0.5899}   \\ \hline
\multicolumn{1}{|l|}{Deterministic QCBO}                         & \multicolumn{1}{l|}{0.0000}   & \multicolumn{1}{l|}{0.9940}   & \multicolumn{1}{l|}{0.9704}   \\ \hline\hline       
\end{tabular}
\label{tbl:1}
\end{table}

Moving on to \emph{S2)}, the goal here is to solve a deterministic QCBO (constrained MaxCut) via VQEC. We do so using the average QCBO and the deterministic QCBO, both solved repeatedly 8 times for different $S$. The circuit depth was kept to $d=3$. Step sizes of VQEC while solving the average QCBO were kept the same as in \emph{S1)}. For the deterministic QCBO, step sizes of primal/dual updates were selected as $\mu^k_{\btheta}=12/(k+10)$ and $\mu^k_{\blambda}=4/(k+15)$, while two additional step sizes of perturbed updates were set to $\nu_{\btheta}=1$ and $\nu_{\blambda}=1.5$. Recall that the two problems differ in how vectors $\bef_m$ are computed. The question is whether the binary vectors $\bb$ drawn from the obtained $\bp(\btheta)$ solve the original deterministic QCBO. The constrained MaxCut has two optimal solutions. We measure the probability of success by summing up the two entries of $\bp(\btheta)$ associated with the two optimal solutions. For each $S$, we define as the final probability of success the worst-case probability of success across the $8$ independent runs. After training the VQC using $S\in\{1,25,50\}$ to get $\btheta$, the obtained $\bp(\btheta)$ was read out using the \texttt{statevector\_ simulator}. Table~\ref{tbl:1} reports these probabilities. For $S=1$, the probability of success of both problems is $0$, which is justifiable as the method has not converged. For $S=25$ and $50$, the variational deterministic QCBO achieved a very good probability of success compared to the variational stochastic QCBO. In other words, the former can serve as an excellent heuristic to provide solutions to QCBOs with high probability.

\begin{figure}[t]
  \centering
  \includegraphics[width=1\linewidth]{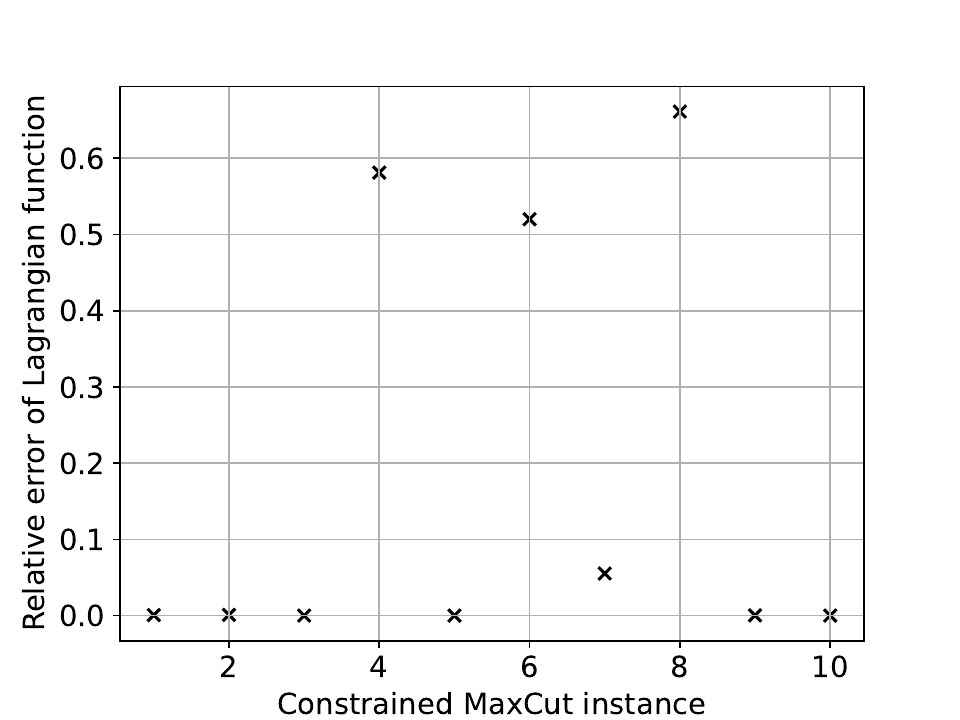}
  \caption{Relative error of the Lagrangian function after convergence while solving the deterministic QCBO for $S=50$ over $10$ constrained MaxCut instances. The step sizes of VQEC and the simulator seed were fixed across instances.}
  \label{fig:QCBO_over_instances}
\end{figure}

For a more thorough evaluation of VQEC in solving the deterministic QCBO, we solved \emph{S2)} over $10$ different constrained MaxCut instances. The optimal value of the Lagrangian function in~\eqref{eq:LagrangianP} found by VQEC was compared to the exact one in~\eqref{eq:Lagrangian} obtained by Gurobi. The number of measurement shots was fixed to $S=50$, and the circuit depth was set to $d=3$. Step sizes of VQEC were set as in the previous test. Figure~\ref{fig:QCBO_over_instances} shows the relative error in the Lagrangian function value $|(\mcL(\btheta^t;\blambda^t)-\mcL^*)/\mcL^*|$ over $10$ different constrained MaxCut instances, where $\mcL^*$ is the exact value of the Lagrangian function found by Gurobi. Although VQEC managed to find optimal or near-optimal solutions for several instances, there exist instances $\{4,6,8\}$ with large relative errors. 

As a sanity check, we repeated the following numerical tests over instances of the constrained MaxCut problem. We solved the related deterministic QCBO by Gurobi and found the minimizer. We then initialized $\btheta$ so that the VQC output coincides with the minimizer. Such initialization can be performed using the procedure described in the proof of Lemma~\ref{le:corners} found in the Appendix. We consequently ran PPD and observed that VQEC did not drift away from the optimal $\btheta$.

\begin{figure}[t]
  \centering
  \includegraphics[width=1.0\linewidth]{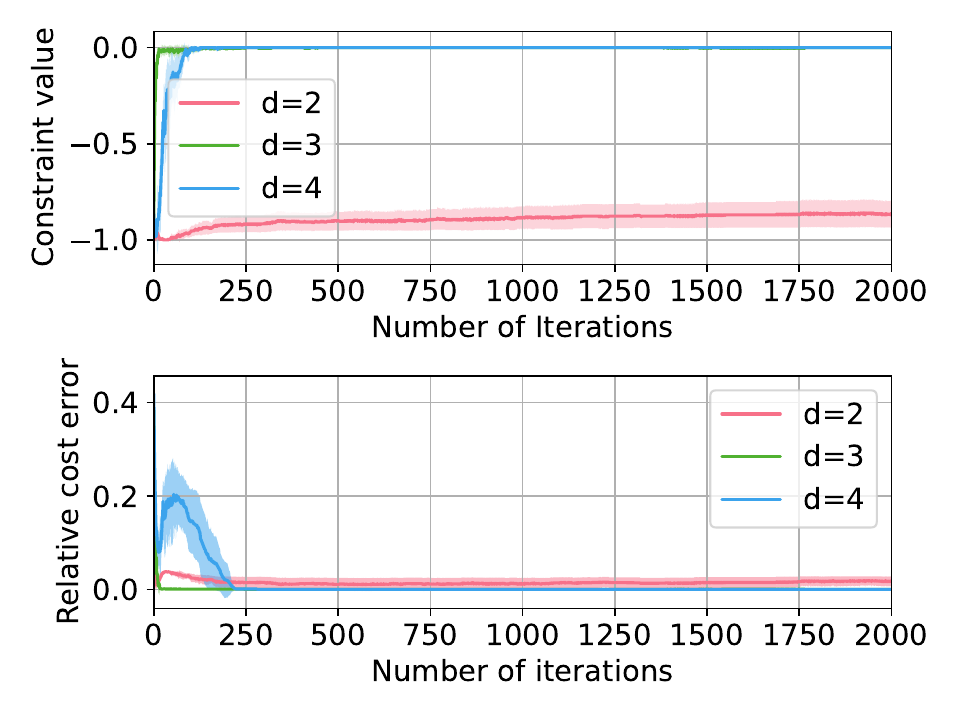}
  \caption{Convergence of constraint function value $F_1(\btheta^t)$ (top) and relative cost error $|(P_\theta^t-P^*)/P^*|$ (bottom) for PPD over the number of PPD iterations while solving \emph{S2)} for $S=50$ and varying values of circuit depths $d$. The plots display confidence intervals within one standard deviation around the mean per iteration, computed over 8 runs.}
  \label{fig:change_D}
\end{figure}

To assess the effect of the circuit depth (dimension of $\btheta$) on the performance of VQEC, we solved the variational deterministic QCBO under \emph{S2)} using circuit depths $d \in \{2, 3, 4\}$. The number of measurement shots was fixed to $S=50$ and each test was repeated 8 times. Figure~\ref{fig:change_D} illustrates the convergence of the constraint value and the relative cost error obtained by PPD for different values of $d$. Clearly, the constraint value and the cost error obtained with $d=2$ deviate substantially from the ideal zero lines. On the contrary, the VQCs with $d=3$ and $d=4$ were able to solve the problem to optimality by increasing the expressibility of the VQC. This observation agrees with recent theoretical findings on the expressibility of VQCs~\cite{schuld2021effect}. It is important to remark that $d=4$ seems to be an over-parameterization of the problem and takes longer to converge. Similar observations were made in~\cite{sim2019}. It is also worth stressing that even for $d=4$, the total number of parameters $P=dn=56$ is much smaller than the size of the original primal LP, that is $N=2^{14}=16,384$. This test signifies that low-depth VQCs might be suboptimal, whereas deeper VQCs may take longer to converge.

\begin{figure}[t]
  \centering
  \includegraphics[width=1\linewidth]{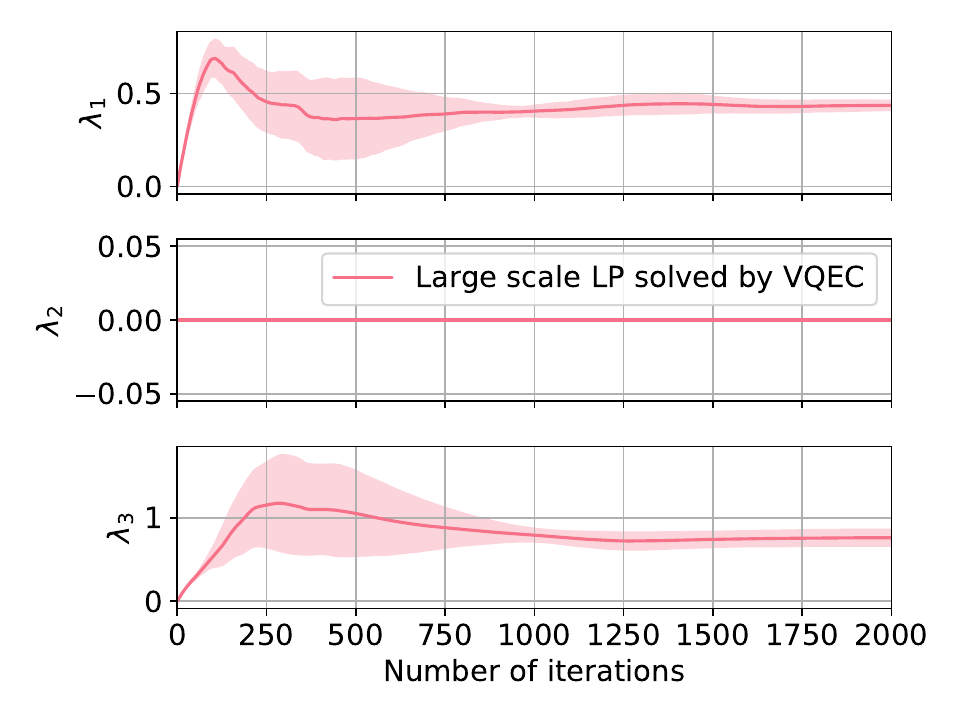}
  \caption{Convergence of entries of the dual variable $\blambda$ over the number of primal/dual iterations under setup \emph{S3)}. The VQEC was repeated 8 times. The plots display confidence intervals within one standard deviation around the mean per iteration, computed over 8 runs.}
  \label{fig:dual_LP}
\end{figure}

\begin{figure}[t]
  \centering
  \includegraphics[width=1\linewidth]{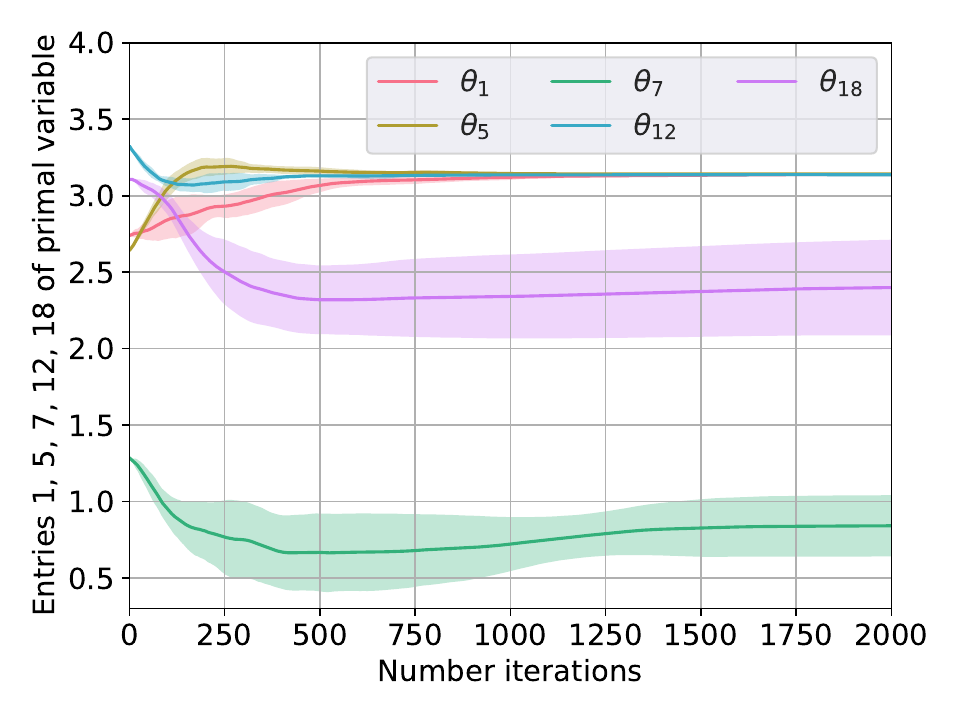}
  \caption{Convergence of entries 1, 5, 7, 12, 18 of the primal variable $\btheta$ over the number of primal/dual iterations under setup \emph{S3)}. The VQEC was repeated 8 times. The plots display confidence intervals within one standard deviation around the mean per iteration, computed over 8 runs.}
  \label{fig:primal_LP}
\end{figure}

\begin{figure}[t]
  \centering
  \includegraphics[width=1\linewidth]{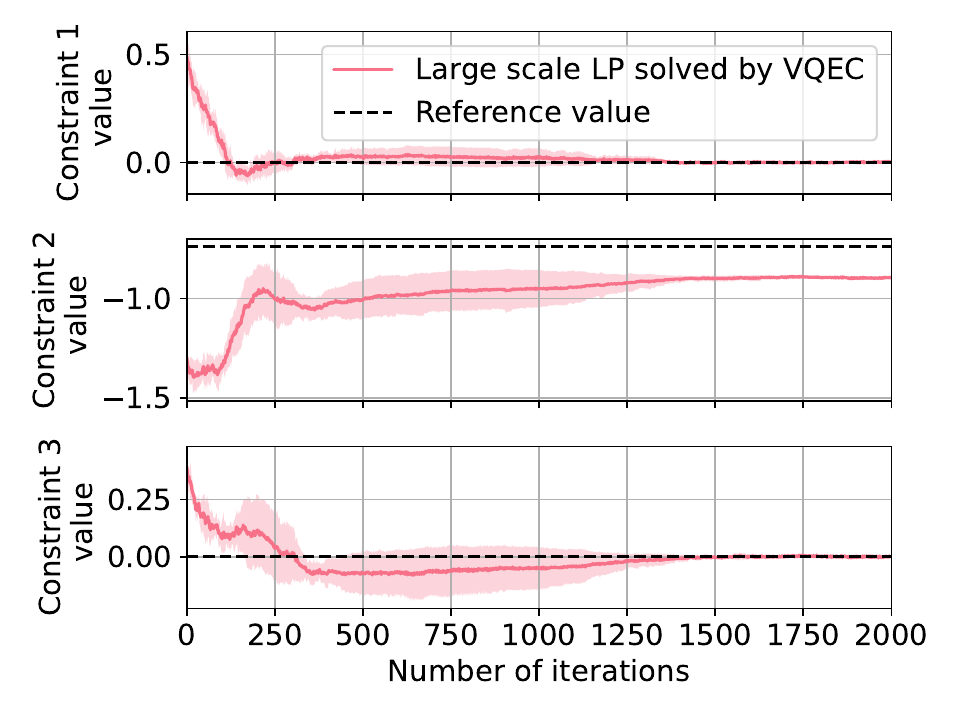}
  \caption{Comparing the constraint values attained by the VQEC and the reference values solved exactly by Gurobi under setup \emph{S3)}. The VQEC was repeated 8 times. The plots display confidence intervals within one standard deviation around the mean per iteration, computed over 8 runs.}
  \label{fig:constraint_LP}
\end{figure}

\begin{figure}[t]
  \centering
  \vspace*{-2em}
  \includegraphics[width=1\linewidth]{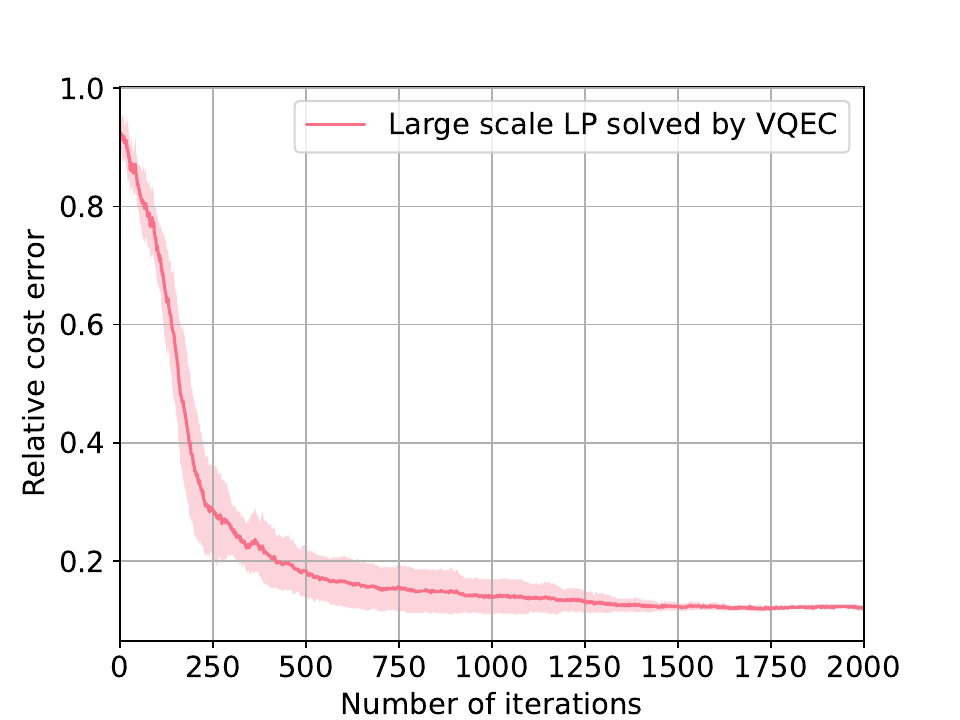}
  \caption{Convergence of the relative cost error under setup \emph{S3)}. The VQEC was repeated 8 times. The plots display confidence intervals within one standard deviation around the mean per iteration, computed over 8 runs.}
  \label{fig:cost_LP}
\end{figure}

Under setup~\emph{S3)}, we evaluated VQEC in solving the large-scale LP in~\eqref{eq:qcbolp}. The step sizes of VQEC were set as $\mu^k_{\btheta}=\mu^k_{\blambda}=0.02 \times 0.999^k$ and $\nu_{\btheta}=\nu_{\blambda}=3$. The number of measurement shots $S$ was set to $150$, and the circuit depth $d$ was set to $3$. The need to increase $S$ for this setup might be caused by the randomly generated values of $\{\bef_m\}_{m=0}^M$, which did not correspond to quadratic functions anymore. VQEC was run 8 times with different values of the simulator seed. The confidence intervals within one standard deviation around the mean per iteration are displayed in Figure~\ref{fig:dual_LP}. Dual variables converged after 2,000 iterations. While $\lambda_2$ converged exactly to 0 for all runs, $\lambda_1$ and $\lambda_3$ varied slightly for different runs. Five randomly chosen entries of the primal variable are shown in Fig.~\ref{fig:primal_LP}. All entries converged after 2,000 iterations. While entries $\theta_{1}$, $\theta_5$, and $\theta_{12}$ converged to roughly the same values over 8 runs, entries $\theta_{7}$ and $\theta_{18}$ varied quite significantly for different runs. To validate the feasibility of the obtained solution, the large-scale LP in~\eqref{eq:qcbolp} was also solved exactly over the exponentially large vector $\bp$ by Gurobi. Figure~\ref{fig:constraint_LP} compares the constraint values found by the proposed method and the reference values by Gurobi. As shown in Fig.~\ref{fig:constraint_LP}, the solution obtained by VQEC satisfied all three constraints. The relative cost error $|(P_\theta^t-P^*)/P^*|$ is shown in Fig.~\ref{fig:cost_LP}, where $P^*$ is the exact cost value obtained by Gurobi. As illustrated in Fig.~\ref{fig:cost_LP}, the relative cost error attained by VQEC converged to around $10\%$ after 2000 iterations. The variance of the relative cost error over 8 runs is very low after convergence. This corroborates that VQEC can generate near-optimal solutions for the large-scale LP in~\eqref{eq:qcbolp}.

\begin{figure}[t]
  \centering
  \includegraphics[width=.9\linewidth]{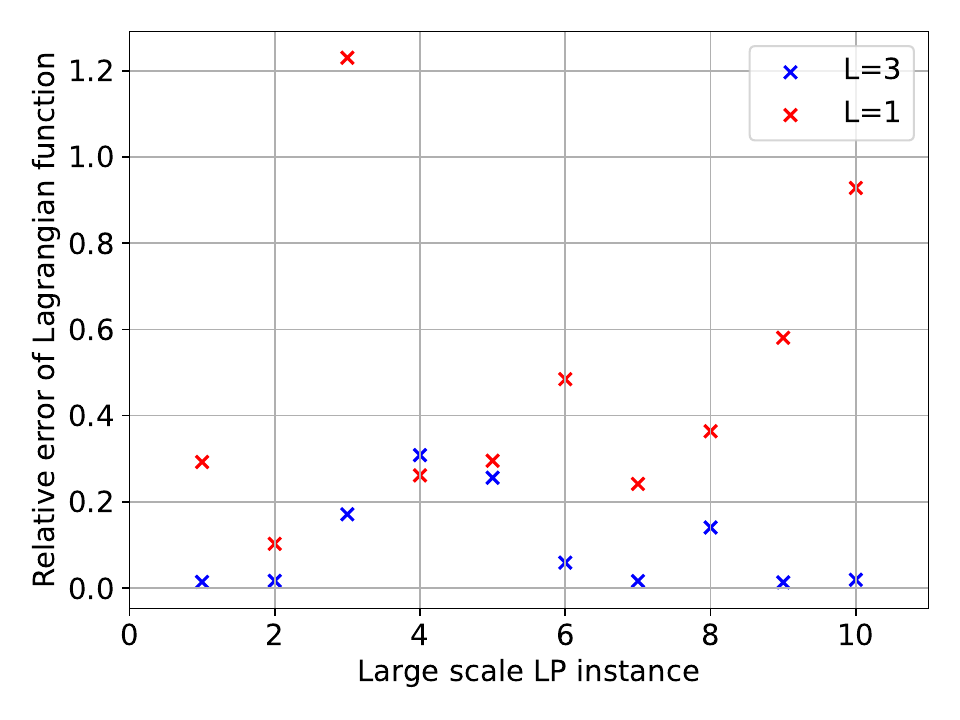}
  \caption{Relative error of the Lagrangian function after $500$ iterations while solving \emph{S3)} for $S=150$ over $10$ large-scale LP instances. The step sizes of VQEC and the simulator seed were fixed across instances.}
  \label{fig:LargeLP_over_instances}
\end{figure}

The optimality gap observed in the previous test reveals that the frequency spectrum of the employed VQC might not be rich enough~\cite{schuld2021effect}. To expand the frequency spectrum, we repeated each VQC parameter of the previous test three times and investigated its effect on the performance of VQEC. Setup~\emph{S3)} was solved over $10$ different large-scale LP instances. The circuit depth $d$ and number of measurement shots $S$ were kept as in the previous test. The step sizes of primal/dual updates were chosen as $\mu^k_{\btheta}=0.05 \times 0.999^k$, $\mu^k_{\blambda}=0.002 \times 0.999^k$, and $\nu_{\btheta}=\nu_{\blambda}=1$. These step sizes and the simulator seed were fixed across the $10$ instances. Figure~\ref{fig:LargeLP_over_instances} compares the relative errors of the Lagrangian function of VQEC with $L=3$ and $L=1$, where $L$ is the number of repeated VQC. The method was deemed to have converged after $500$ iterations. As shown in Fig.~\ref{fig:LargeLP_over_instances}, for $9$ instances, VQEC with $L=3$ performed much better than the one with $L1$. Among those $9$ cases, $8$ instances attained relative errors below $20\%$. This test shows that repeating parameters enhances the performance of VQEC.

\section{Conclusions}\label{sec:conclusions}
This work has developed, analyzed, and evaluated a novel algorithm for handling optimization problems with constraints using a variational quantum approach. The proposed VQEC deals with constraints in the dual domain of the parameterized variational quantum problem. Primal and dual variables are iteratively updated on a classical computer via stochastic PPD iterates. Compared to VQE, VQEC entails insignificant computational overhead as the cost and constraint observables are measured simultaneously. VQEC has been applied to diverse problems defined over diagonal observables, including binary programs with constraints, finding optimal stochastic policies to draw binary vectors satisfying average and chance constraints, learning probability mass functions, and solving large-scale LPs over the probability simplex. The possible performance degradation by solving a problem in its quantum variational form \emph{vis-à-vis} its original form has been characterized. Extensive numerical tests using IBM's quantum simulator have corroborated that: \emph{i)} The PPD method performs remarkably better than the PD method with only one additional measurement shot per primal/dual iteration; \emph{ii)} VQEC with finite numbers of measurement shots can yield meaningful solutions to QCBOs, stochastic QCBOs, and large-scale LPs over the probability simplex; and \emph{iii)} The solution quality and/or convergence rate of VQEC can be affected by under/over-parameterized VQCs. 

In a nutshell, VQEC provides a principled solution to the practically relevant task of incorporating constraints into variational quantum approaches. It can thus patently expand the application domain of NISQ computers. Capitalizing on the promising results of this work, we are currently working toward several exciting research directions, such as: \emph{d1)} Coping with non-diagonal observables is highly desirable as it clearly broadens the applicability of VQEC from binary and linear programs to conic programs, including quadratically constrained quadratic programs (QCQPs) and semidefinite programs (SDPs). Dealing with mixed-integer programs involving binary and continuous variables is on current focus too; \emph{d2)}
The VQCs considered thus far have been confined to be parameterized solely by $\btheta$ to produce state $\ket{\bx(\btheta)}=\bS(\btheta)\ket{0}_n$. Nonetheless, in a quantum machine learning setting, a VQC may also encode \emph{data (features)} $\{\bz_t\}_{t=1}^T$ in the form of a data-embedding mechanism as in $\bU(\btheta;\bz_t)=\bS(\btheta)\bW(\bz_t)$ or other forms. Measuring the quantum state $\ket{\bx(\btheta;\bz_t)}=\bU(\btheta;\bz_t)\ket{0}_n$ and applying VQEC to proper loss functions, can tackle pertinent (un)supervised quantum machine learning tasks; \emph{d3)} In light of the performance analysis of Section~\ref{sec:analysis}, it is vital to further investigate Assumption~\ref{as:epsilon} on the accuracy with which VQCs can approximate PMFs and other optimization objects; \emph{d4)} Extending our performance analysis to other optimization classes and contrasting VQEC to stochastic classical counterparts.


\begin{acknowledgments}
This work was supported in part by the U.S. National Science Foundation under Grant 1751085.
\end{acknowledgments}
\bibliography{reference}

\appendix
\section{Proof of Theorem~\ref{th:zdg}}
To establish zero duality gap, it suffices to show that the perturbation function associated with the parameterized primal problem is convex~\cite{Rockafellar}. This perturbation function is defined by modifying the right-hand side of the linear inequality constraints of \eqref{eq:primalP} as
\begin{align}\label{eq:primalPperturbed}
P_{\theta}=\min_{\btheta}~&~\bef_0^\top \bp(\btheta)\\
\mathrm{s.to}~&~\bF^\top\bp(\btheta)\leq \bdelta\nonumber
\end{align}
where $\bdelta> -\bs_{\theta}$ with $\bs_{\theta}$ from Assumption~\ref{as:strictlyfeasible}. Consider two points $\bdelta_1$ and $\bdelta_2$. Suppose solving \eqref{eq:primalPperturbed} for $\bdelta_1$ yields the minimizer $\btheta_1$ generating $\bp_1=\bp(\btheta_1)$. Likewise, solving \eqref{eq:primalPperturbed} for $\bdelta_2$ yields the minimizer $\btheta_2$ generating $\bp_2=\bp(\btheta_2)$. Under Assumption~\ref{as:convexP}, any vector $\bp_{\alpha}=\alpha\bp_1+(1-\alpha)\bp_2$ for $\alpha\in[0,1]$ belongs to the set $\mcP_{\theta}$ of admissible PMF vectors. That means that there exists a $\btheta_{\alpha}$ for which $\bp_{\alpha}=\bp(\btheta_{\alpha})$. Vector $\bp_{\alpha}$ is feasible for the perturbed problem if the latter is perturbed by $\bdelta_{\alpha}=\alpha\bdelta_1+(1-\alpha)\bdelta_2$ because
\[\bF^\top\bp_{\alpha}=\alpha\bF^\top\bp_1+(1-\alpha)\bF^\top\bp_2\leq \alpha\bdelta_1+(1-\alpha)\bdelta_2=\bdelta_{\alpha}.\]
Since $\bp_{\alpha}$ is feasible for the perturbed problem for $\bdelta_{\alpha}$, it provides an upper bound for the perturbation function $P_{\theta}(\bdelta_{\alpha})$ as
\begin{align*}
P_{\theta}(\bdelta_{\alpha}) &\leq \bef_0^\top\bp_{\alpha}=\alpha\bef_0^\top\bp_1+(1-\alpha)\bef_0^\top\bp_2\\
&=\alpha P_{\theta}(\bdelta_1)+(1-\alpha)P_{\theta}(\bdelta_2).
\end{align*}
This establishes that $P_{\theta}(\bdelta)$ is a convex function of $\bdelta$, which in turn implies that the parameterized problem has zero duality gap.

\section{Proof of Lemma~\ref{le:corners}}
Consider first a single layer of the VQC shown in Fig.~\ref{fig:two_local}. Suppose the input to this layer is $\ket{0}_n$. The layer consists of a parameterized block and an entanglement block. The parameterized block applies gate
\[\bR_Y(2\theta_p)= \exp\left(-i\theta_p\bY\right)=
	\begin{bmatrix}
		\cos(\theta_p)& -\sin(\theta_p)\\
		\sin(\theta_p)& \cos(\theta_p)
	\end{bmatrix}\]
on qubit $p$ for $p=1,\ldots,n$, where $\bY$ is the matrix representing the Pauli $Y$ gate. If $\ket{\bpsi}$ denotes the state after the parameterized block, its $p$-th qubit is
\[\ket{\bpsi}_p = \bR_Y(2\theta_p)\ket{0} = \cos(\theta_p)\ket{0}+\sin(\theta_p)\ket{1},~~p=1:n.\]

The entanglement block includes controlled-Z (CZ) gates between all pairs of qubits. Since the roles of the control and target qubits are interchangeable for the CZ gate, we designate control and target qubits as shown in Fig.~\ref{fig:two_local} without loss of generality. Let $\ket{\bphi}$ be the state after the entanglement block. As $\ket{\bpsi}_1$ is not subjected to any control gate, qubit $1$ of $\ket{\bphi}$ is
\begin{equation}\label{eq:phi1}
\ket{\bphi}_1=\ket{\bpsi}_1=\cos(\theta_1)\ket{0}+\sin(\theta_1)\ket{1}.
\end{equation}
By drawing $\theta_1$ from $\{0,\pi/2\}$, qubit $\ket{\bphi}_1$ can take any value in $\{\ket{0},\ket{1}\}$.

The entanglement block shown in Fig.~\ref{fig:two_local} consists of $(n-1)$ sub-blocks. The first sub-block comprises $(n-1)$ CZ gates, in which qubit $1$ controls qubits $2$ to $n$. The target qubits after this sub-block are denoted as $\ket{\bphi}^1_p$ for $p=2,\ldots,n$. The second sub-block comprises $(n-2)$ CZ gates, in which qubit $2$ controls qubits $3$ to $n$. The target qubits after this sub-block are denoted as $\ket{\bphi}^2_p$ for $p=3,\ldots,n$. Subsequent sub-blocks expand similarly up to sub-block $(n-1)$. 

Consider the output of the first sub-block. If $\ket{\bphi}_1=\ket{0}$, then $\ket{\bphi}^1_p=\ket{\bpsi}_p=\cos(\theta_p)\ket{0}+\sin(\theta_p)\ket{1}$. If $\ket{\bphi}_1=\ket{1}$, then $\ket{\bphi}^1_p=\bZ\ket{\bpsi}_p=\cos(\theta_p)\ket{0}-\sin(\theta_p)\ket{1}$. Since $\ket{\bphi}_1=\ket{0}$ implies $\sin(\theta_1)=0$, and $\ket{\bphi}_1=\ket{1}$ implies $\sin(\theta_1)=1$, the output of the first sub-block can be compactly expressed as
\[\ket{\bphi}^1_p=\cos(\theta_p)\ket{0}+(-1)^{\sin(\theta_1)}\sin(\theta_p)\ket{1},~~ p=2:n.\]
Because qubit 2 is not controlled by any more qubits, its state can be finalized here as
\begin{equation}\label{eq:bphi_2}
\ket{\bphi}_2=\ket{\bphi}^1_2=\cos(\theta_2)\ket{0}+(-1)^{\sin(\theta_1)}\sin(\theta_2)\ket{1}.
\end{equation}
It is now easy to verify that if we want to set $\ket{\bphi}_2=\ket{0}$, we can simply set $\theta_2=0$. Otherwise, that is to set $\ket{\bphi}_2=\ket{1}$, parameter $\theta_2$ is selected between $\{\pi/2,3\pi/2\}$ depending on the value of $\theta_1$. Specifically, if $\theta_1=0$, set $\theta_2=\pi/2$; and if $\theta_1=\pi/2$, set $\theta_2=3\pi/2$ to get the proper sign. 

Consider now the second sub-block, where CZ gates are controlled by $\ket{\bphi}_2$. If $\ket{\bphi}_2=\ket{0}$, then $\ket{\bphi}^2_p=\ket{\bphi}^1_p=\cos(\theta_p)\ket{0}+(-1)^{\sin(\theta_1)}\sin(\theta_p)\ket{1}$. If $\ket{\bphi}_2=\ket{1}$, then $\ket{\bphi}^2_p=\bZ\ket{\bphi}^1_p=\cos(\theta_p)\ket{0}-(-1)^{\sin(\theta_1)}\sin(\theta_p)\ket{1}$. The two cases can be compactly expressed as
\[\ket{\bphi}^2_p=\cos(\theta_p)\ket{0}+(-1)^{\sin(\theta_1)+\sin(\theta_2)}\sin(\theta_p)\ket{1}\]
for $p=3:n$. This is because if $\ket{\bphi}_2=\ket{0}$, equation~\eqref{eq:bphi_2} implies that $\cos(\theta_2)=1$, and thus, $\sin(\theta_2)=0$ and $(-1)^{\sin(\theta_2)}=1$. Otherwise, that is if $\ket{\bphi}_2=\ket{1}$, equation~\eqref{eq:bphi_2} yields $\cos(\theta_2)=0$, and thus, $\sin(\theta_2)=\pm1$ and $(-1)^{\sin(\theta_2)}=-1$. Because qubit 3 is not controlled by any more qubits, its state can be finalized here as 
\[\ket{\bphi}_3=\ket{\bphi}^2_3=\cos(\theta_3)\ket{0}+(-1)^{\sin(\theta_1)+\sin(\theta_2)}\sin(\theta_3)\ket{1}.\]

The previous argument carries along subsequent sub-blocks. Therefore, qubit $p$ of $\ket{\bphi}$ can be expressed as
\begin{equation}\label{eq:bphi_p}
\ket{\bphi}_p=\cos(\theta_p)\ket{0}+(-1)^{\sum_{i=1}^{p-1}\sin(\theta_i)}\sin(\theta_p)\ket{1}
\end{equation}
for $p=1:n$. The formula dictates how to set parameter $\theta_p$ so that qubit $p$ takes a particular binary value. The process transitions from the first to the last qubit. For qubit $p$ to be set to $\ket{0}$, simply set $\theta_p=0$. For qubit $p$ to be set to $\ket{1}$, select $\theta_p=\pi/2$ if ${\sum_{i=1}^{p-1}\sin(\theta_i)}$ is even, or $\theta_p=3\pi/2$ if ${\sum_{i=1}^{p-1}\sin(\theta_i)}$ is odd. This shows that state $\ket{\bphi}$ can span all canonical vectors in $\mathbb{R}^N$. 

So far, we have considered a single layer of the VQC. If there are $d$ layers, we can set the parameters of layers $1$ to $(d-1)$ to zero, and then apply the established claim for the single layer only to the last layer. Hence, when layer $1$ is fed with $\ket{0}_n$, its output remains $\ket{0}_n$ and is fed as input to layer $2$. The last layer is eventually fed with $\ket{0}_n$, and can thus, be treated as a single layer. 

The previous analysis holds for the linear and circular entanglements too. Consider again a single layer. For the linear entanglement, CZ gates are implemented between successive qubits. Accordingly, the first qubit of $\ket{\bpsi}$ again is not subjected to any control gate, and so \eqref{eq:phi1} still applies. As each qubit after the first one is controlled by its previous qubit, it is easy to see that
\begin{equation}\label{eq:bphi_p_linear}
\ket{\bphi}_p=\cos(\theta_p)\ket{0}+(-1)^{\sin(\theta_{p-1})}\sin(\theta_p)\ket{1}
\end{equation}
for $p=2:n$.
Therefore, by properly sampling $\{\theta_p\}_{p=1}^n$ from $\{0,\pi/2,3\pi/2\}$, state $\ket{\bphi}$ can be made to take the value of any canonical vector in $\mathbb{R}^N$.

Compared to the linear, the circular entanglement differs only in the first qubit, which becomes $\ket{\bphi}_1=\cos(\theta_1)\ket{0}+(-1)^{\sin(\theta_n)}\sin(\theta_1)\ket{1}$. Albeit the recursion is different, each qubit of $\ket{\bphi}$ can again evaluate to either $\ket{0}$ or $\ket{1}$. Specifically, for qubit $p$ to be selected to $\ket{0}$, we set $\theta_p=0$. For qubit $p$ to be selected to $\ket{1}$, we set $\theta_p=\pi/2$ if its previous qubit is $\ket{0}$; and $\theta_p=3\pi/2$, if its previous qubit is $\ket{1}$.

\section{Proof of Theorem~\ref{th:degradation}}
We commence with the lower bound on $D_{\theta}^*$. If $\blambda^*$ and $\blambda_{\theta}^*$ are the maximizers of the dual problems in \eqref{eq:dual*} and \eqref{eq:dual*P}, respectively, we will prove that
\begin{equation}\label{eq:sequence}
D^*=D(\blambda^*) \leq D_{\theta}(\blambda^*) \leq D_{\theta}(\blambda_{\theta}^*)=D_{\theta}^*.
\end{equation}
The equalities in \eqref{eq:sequence} hold obviously by definition. To show the first inequality in \eqref{eq:sequence}, note that the dual functions $D(\blambda)$ and $D_{\theta}(\blambda)$ are both defined as the result of the minimization problems defined in \eqref{eq:dual} and \eqref{eq:dualP}. Because \eqref{eq:dualP} is a restriction of \eqref{eq:dual}, it implies that $D(\blambda) \leq D_{\theta}(\blambda)$ for all $\blambda\geq \bzero$. Plugging $\blambda^*$ in the previous inequality provides the first inequality in \eqref{eq:sequence}. The second inequality in \eqref{eq:sequence} holds simply because $\blambda_{\theta}^*$ maximizes the dual function $D_{\theta}(\blambda)$ over all $\blambda\geq \bzero$.

We proceed with the upper bound on $D_{\theta}^*$ in \eqref{eq:degradation}. Upon combining \eqref{eq:dualP}--\eqref{eq:dual*P}, we can express
\[D_{\theta}^*=\max_{\blambda\geq \bzero}\min_{\btheta}~\mcL_{\theta}(\btheta;\blambda)\]
Using the definition of the two Lagrangian functions in \eqref{eq:Lagrangian} and \eqref{eq:LagrangianP}, we can write
\begin{align}\label{eq:bound1}
\mcL_{\theta}(\btheta;\blambda)&=(\bef_0+\bF\blambda)^\top\bp+  (\bef_0+\bF\blambda)^\top(\bp(\btheta)-\bp)\nonumber\\
&=\mcL(\bp;\blambda) + (\bef_0+\bF\blambda)^\top(\bp(\btheta)-\bp)\nonumber\\
&\leq \mcL(\bp;\blambda) + \|\bef_0+\bF\blambda\|_1\cdot \|\bp(\btheta)-\bp\|_{\infty}
\end{align}
where the inequality follows from H\"{o}lder's inequality.

The $\ell_1$-norm in \eqref{eq:bound1} can be upper bounded using the triangle and matrix inequalities as
\begin{equation}\label{eq:bound2}
\|\bef_0+\bF\blambda\|_1\leq \|\bef_0\|_1+\|\bF\|_1\cdot\|\blambda\|_1\leq \|\bef_0\|_1+L\|\blambda\|_1
\end{equation}
with $L$ defined in \eqref{eq:L}. Plugging \eqref{eq:bound2} into \eqref{eq:bound1} provides
\begin{equation*}
\mcL_{\theta}(\btheta;\blambda)\leq \mcL(\bp;\blambda) + \left(\|\bef_0\|_1+L\|\blambda\|_1\right)\cdot \|\bp(\btheta)-\bp\|_{\infty}.
\end{equation*}
The last inequality holds for all $\btheta$. If we minimize both sides over $\btheta$, the direction of the inequality remains\footnote{It is easy to show that if $g_1(\btheta)\leq g_2(\btheta)$ for all $\btheta$, then $\min_{\btheta}g_1(\btheta)\leq \min_{\btheta}g_2(\btheta)$ and $\max_{\btheta}g_1(\btheta)\leq \max_{\btheta}g_2(\btheta)$.}:
\begin{align}\label{eq:bound4}
&\min_{\btheta}\mcL_{\theta}(\btheta;\blambda)\leq \mcL(\bp;\blambda)\nonumber\\
&\quad\quad\quad\quad +  \left(\|\bef_0\|_1+L\|\blambda\|_1\right) \cdot \min_{\btheta}\|\bp(\btheta)-\bp\|_{\infty}.
\end{align}

The inequality in~\eqref{eq:bound4} holds for all $\bp\in\mcP$. Consider a particular $\bp\in\mcP$. By Assumption~\ref{as:epsilon}, there exists $\btheta_0$ for which $\|\bp-\bp(\btheta_0)\|_{\infty}\leq \epsilon$ for this particular $\bp$. The parameter vector $\btheta_0$ provides the upper bound 
\begin{equation}\label{eq:bound5}
\min_{\btheta}\|\bp(\btheta)-\bp\|_{\infty}\leq \|\bp(\btheta_0)-\bp\|_{\infty}\leq \epsilon
\end{equation}
which holds for all $\bp\in\mcP$. Plugging \eqref{eq:bound5} into \eqref{eq:bound4} yields
\begin{equation}\label{eq:bound6}
\min_{\btheta}~\mcL_{\theta}(\btheta;\blambda)\leq \mcL(\bp;\blambda)+  \epsilon\|\bef_0\|_1+\epsilon L\|\blambda\|_1.
\end{equation}
Minimizing both sides of \eqref{eq:bound6} over $\bp\in\mcP$ provides
\begin{equation}\label{eq:bound7}
\min_{\btheta}\mcL_{\theta}(\btheta;\blambda)\leq \min_{\bp\in\mcP}~\mcL(\bp;\blambda) +\epsilon\|\bef_0\|_1+\epsilon L\|\blambda\|_1.
\end{equation}
The last inequality in~\eqref{eq:bound7} holds for all $\blambda\geq \bzero$. If we maximize both sides over $\blambda\geq \bzero$, the direction of the inequality remains and yields:
\begin{align}\label{eq:bound8}
&D_{\theta}^*=\max_{\blambda\geq \bzero}\min_{\btheta}~\mcL_{\theta}(\btheta;\blambda)\nonumber\\
&\quad \quad\quad\quad\leq \epsilon\|\bef_0\|_1+\max_{\blambda\geq \bzero}\min_{\bp\in\mcP}~\mcL(\bp;\blambda) +\epsilon L\|\blambda\|_1
\end{align}

The second summand in the right-hand side of \eqref{eq:bound8} seems to be related to $P^*=D^*=\max_{\blambda\geq \bzero}\min_{\bp\in\mcP}~\mcL(\bp;\blambda)$. Unfortunately, the maximization involves the additional term $\epsilon L \|\blambda\|_1$. Interestingly, this second summand relates to the perturbed primal problem introduced in \eqref{eq:primalperturbed}. More specifically, the Lagrangian function of \eqref{eq:primalperturbed} is defined as
\[\tilde{\mcL}(\bp;\blambda):=(\bef_0+\bF\blambda)^\top\bp +\epsilon L \|\blambda\|_1=\mcL(\bp;\blambda)+\epsilon L \|\blambda\|_1\]
where we have used the property that $\blambda^\top\bone=\|\blambda\|_1$ since $\blambda\geq \bzero$. Under Assumption~\ref{as:perturbedprimal}, the perturbed primal problem is feasible, and thus, strong duality holds. Therefore, the optimal cost of the perturbed problem in \eqref{eq:primalperturbed} equals 
\[\tilde{P}^*=\max_{\blambda\geq \bzero}\min_{\bp\in\mcP}\tilde{\mcL}(\bp;\blambda)=\max_{\blambda\geq \bzero}\min_{\bp\in\mcP}~\mcL(\bp;\blambda) +\epsilon L\|\blambda\|_1.\]
The second summand on the right-hand side of \eqref{eq:bound8} coincides with $\tilde{P}^*$. Because the original primal LP is convex, the optimal cost of the perturbed problem is known to satisfy~\cite[Sec.~5.6.2]{BoVa04}:
\[P^*\geq  \tilde{P}^* -\epsilon L\|\tblambda\|_1\]
where $\tblambda$ is the vector of optimal Lagrange multipliers for \eqref{eq:primalperturbed}. The last inequality provides an upper bound on $\tilde{P}^*$. Plugging this bound into \eqref{eq:bound8} gives
\begin{equation}\label{eq:bound9}
D_{\theta}^*\leq P^* + \epsilon\|\bef_0\|_1+\epsilon L\|\tblambda\|_1    
\end{equation}
and proves the upper bound on $D_{\theta}^*$ in \eqref{eq:degradation}.

To bound $\|\tblambda\|_1$, a standard trick can be adopted; see~\cite[Ex.~5.3.1]{Be99} and \cite{eisen19}. Under Assumption~\ref{as:perturbedprimal}, the perturbed primal problem in \eqref{eq:primalperturbed} is strictly feasible. Multiplying both sides of \eqref{eq:strictfeasibleperturbedprimal} by $\tblambda\geq \bzero$ and summing up gives
\begin{equation}\label{eq:bound10}
\tblambda^\top\bF^\top\hbp\leq -\epsilon L \|\tblambda\|_1 - s_0 \|\tblambda\|_1.
\end{equation}
Because problem \eqref{eq:primalperturbed} satisfies strong duality and $\hbp\in\mcP$ is feasible for \eqref{eq:primalperturbed}, it follows that
\begin{align*}
\tilde{P}^*&=\min_{\bp\in\mcP}~(\bef_0+\bF\tblambda)^\top\bp + \epsilon L \|\tblambda\|_1\\
&\leq (\bef_0+\bF\tblambda)^\top\hbp + \epsilon L \|\tblambda\|_1\\
&\leq \bef_0^\top \hbp + \tblambda^\top \bF^\top\hbp+ \epsilon L \|\tblambda\|_1\\
&\leq \bef_0^\top \hbp - s_0\|\tblambda\|_1
\end{align*}
where the last inequality stems from \eqref{eq:bound10}. 

Because the perturbed problem in \eqref{eq:primalperturbed} is a restriction of the original primal LP in \eqref{eq:primal}, we get that
\[P^*\leq\tilde{P}^*\leq\bef_0^\top \hbp - s_0\|\tblambda\|_1\]
from which we obtain the bound in \eqref{eq:degradation2}. This completes the proof of Theorem~\ref{th:degradation}.

\end{document}